%% file: main.tex
\documentclass[journal,10pt,twocolumn,twoside]{IEEEtran}
\pdfoutput=1
\usepackage{amsmath, amssymb}
\usepackage[utf8]{inputenc}
\usepackage[T1]{fontenc}
\usepackage{tikz}
\usepackage{cite}
\usepackage{diagbox}
\usepackage{xcolor}
\ifCLASSOPTIONcompsoc
 \usepackage[subrefformat=parens, labelformat=parens, caption=false,font=normalsize,labelfont=sf,textfont=sf]{subfig}
\else
 \usepackage[subrefformat=parens, labelformat=parens, caption=false,font=footnotesize]{subfig}
\fi

\def\figpath{.}

\input{newcommands}

\DeclareMathOperator{\vect}{vec}

\DeclareMathOperator{\e}{E}

\DeclareMathOperator{\sinc}{sinc}

\DeclareMathOperator{\diag}{diag}
\DeclareMathOperator{\rank}{rank}
\DeclareMathOperator{\reshape}{reshape}
\DeclareMathOperator{\blkdiag}{blkdiag}
\newcommand{\fnorm}[1]{\left\|#1\right\|_F}
\newcommand{\modd}[1]{\left\langle#1\right\rangle}
\newtheorem{theorem}{Theorem}
\newtheorem{lemma}{Lemma}
\newtheorem{corollary}{Corollary}
\newtheorem{definition}{Definition}

\usepackage{makecell}

\begin{document}
%
\title{Matrix Characterization for GFDM: Low Complexity MMSE Receivers and Optimal Filters}

\author{Po-Chih~Chen, ~Borching~Su, ~and~Yenming~Huang

\ignore{

\thanks{M. Shell was with the Department
of Electrical and Computer Engineering, Georgia Institute of Technology, Atlanta,
GA, 30332 USA e-mail: (see http://www.michaelshell.org/contact.html).}
\thanks{J. Doe and J. Doe are with Anonymous University.}
\thanks{Manuscript received April 19, 2005; revised August 26, 2015.}}

}

\ignore{
\markboth{To be submitted to IEEE Trans. on Signal Processing}{Chen and Su: Matrix Characterization for Generalized Frequency Division Multiplexing Systems}
}
%



\maketitle

\begin{abstract}
In this paper, a new matrix-based characterization of generalized-frequency-division-multiplexing (GFDM) transmitter matrices is proposed, as opposed to traditional vector-based characterization with prototype filters.
The characterization facilitates deriving properties of GFDM (transmitter) matrices, including conditions for GFDM matrices being nonsingular and unitary, respectively. 
Using the new characterization, the necessary and sufficient conditions for the existence of a form of low-complexity implementation for a minimum mean square error (MMSE) receiver are derived.
Such an implementation exists under multipath channels if the GFDM transmitter matrix is selected to be unitary. 
For cases where this implementation does not exist, a low-complexity suboptimal MMSE receiver is proposed, with its performance approximating that of an MMSE receiver.
The new characterization also enables derivations of optimal prototype filters in terms of minimizing receiver mean square error (MSE).
They are found to correspond to the use of unitary GFDM matrices under many scenarios.
The use of such optimal filters in GFDM systems does not cause the problem of noise enhancement, thereby demonstrating the same MSE performance as orthogonal frequency division multiplexing.
Moreover, we find that GFDM matrices with a size of power of two are verified to exist in the class of unitary GFDM matrices.
Finally, while the out-of-band (OOB) radiation performance of systems using a unitary GFDM matrix is not optimal in general, it is shown that the OOB radiation can be satisfactorily low if parameters in the new characterization are carefully chosen.

\end{abstract}


%
\IEEEpeerreviewmaketitle

\section{Introduction}
%
%
%
%
\IEEEPARstart{G}{eneralized} frequency division multiplexing (GFDM) \cite{fettweis09}, extensively studied in recent years, is a potential modulation scheme for future wireless communication systems because it features good properties including low out-of-band (OOB) radiation and flexible time-frequency structures to adapt to various application scenarios, such as cognitive radios and low latency applications \cite{michailow14}.
However, some drawbacks for GFDM arise from the non-orthogonality \cite{matthe14a} of the system as a result of using prototype transmit filters \cite{michailow14}. In this study, we address two specific drawbacks: the difficulty in designing low-complexity transceivers, and performance degradation in receiver mean square error (MSE) and symbol error rate (SER) compared to that achieved through orthogonal frequency division multiplexing (OFDM) \cite{bingham90}. 
The severity of the performance degradation depends heavily on the prototype transmit filter that is selected \cite{matthe14a}.

For GFDM systems with a matched filter (MF) receiver \cite{michailow14, matthe14b}, inter-carrier interference (ICI) and inter-symbol interference (ISI) exist. To cancel ICI and ISI, successive interference cancellation (SIC) receivers are employed \cite{datta12a, michailow14, datta11}. 
However, long delays are incurred in the process of interference cancellation. 
In this paper, we focus on zero-forcing (ZF) and linear minimum mean square error (MMSE) receivers \cite{michailow14, matthe14b}, which eliminate ICI and ISI. 
Although the ZF receiver is known for its low-complexity implementation under either additive white Gaussian noise (AWGN) or multipath channels, MMSE receiver implementations with linearithmic complexity, to the best of our knowledge, is known only for the AWGN channels (see recent references \cite{matthe16b, farhang16}). In \cite{matthe15d,matthe16a}, MMSE receivers for multipath channels with reduced complexity were proposed, but they still have at least a quadratic complexity (in terms of numbers of GFDM subsymbols or subcarriers). 
In this paper, we study the feasibility of low-complexity MMSE receivers in presence of multipath channels and propose the first implementation with linearithmic complexity thereof.

In addition, we study the impact of GFDM prototype transmit filters on MSE and SER performance.
In the literature \cite{michailow12b, michailow14, matthe14b, fettweis09, michailow12a, bandari15, matthe14a, alves13, tiwari15, rezazadehReyhani15, matthe15a, matthe15b, matthe15c, sharifian15, li16, farhang16, han16arxiv, gaspar15c, chang16, wang16, gaspar14b, gaspar15b, matthe16b}, many prototype filters, including the raised-cosine (RC), root-raised-cosine (RRC), Xia \cite{xia97}, Dirichlet \cite{matthe14a}, and Gaussian pulses, have been proposed and used for GFDM systems.
These prototype filters are mostly designed to reduce OOB radiation of transmitted signals except that the Dirichlet pulse is claimed to be rate-optimal under the ZF or MMSE receiver over the AWGN channel \cite{han16arxiv}.
However, GFDM systems using all these filters are mostly non-orthogonal (except the Dirichlet pulse) \cite{matthe14a}. In other words, the corresponding GFDM transmitter matrices \cite{michailow14} generally have a greater-than-unity condition number. This creates the noise enhancement effect \cite{matthe14b, michailow12a, han16arxiv}, and GFDM systems using these filters suffer from MSE and SER performance degradation compared to OFDM systems.


This study offers three main contributions:
\subsubsection{New matrix characterization of GFDM transceivers} 
    The modulation process in a GFDM transmitter can be performed by multiplying the data vector by a matrix with a special structure, called a GFDM matrix. A GFDM matrix is commonly characterized by its first column, usually referred to as the \emph{prototype filter} \cite{michailow14}. In some other references \cite{michailow12b, matthe16b}, a GFDM matrix is characterized by the \emph{frequency-domain prototype filter}, i.e., the discrete Fourier transform (DFT) of the prototype filter, which leads to some advanced implementations of GFDM transceivers. In this paper, we propose an alternative means for characterizing GFDM matrices, in which a \emph{characteristic matrix} is used. On the basis of this new characterization, we investigated several properties of GFDM matrices and found that the conditions for some properties of a GFDM matrix (e.g., non-singularity, unitary property) can be expressed very clearly with the new characterization parameters. This characterization also leads to low-complexity transmitter implementations and provides a foundation for the other two contributions, described as follows.

\subsubsection{Low-complexity MMSE receivers under multipath channels} In this paper, we propose a form of low-complexity implementation for an MMSE receiver.
The necessary and sufficient conditions for the existence of such an implementation are derived and clearly expressed in terms of the new matrix characterization parameters.
Particularly, the use of a \emph{unitary} GFDM transmitter matrix is a sufficient condition.
Moreover, for cases where the necessary condition is not satisfied, we also propose a low-complexity suboptimal MMSE receiver whose performance approximates that of an MMSE receiver. This makes GFDM transceivers very practicable even in multipath channels.
The complexity of our proposed implementation is analyzed in detail and compared to existing solutions. We show that  significant complexity reduction can be obtained through the use of our implementation.
\subsubsection{Optimal prototype transmit filters in receiver MSE} In this study, we investigate the optimal prototype transmit filters in terms of minimizing receiver MSEs with both ZF and MMSE receivers under the AWGN channel as well as static and statistical linear time-invariant channels. We find that the optimal GFDM transmitter matrices under most scenarios are unitary GFDM matrices and do not suffer from the noise enhancement effect. Besides, we identify several unitary GFDM matrices that achieve sufficiently favorable OOB radiation performance for practical applications.


The remainder of this paper is structured as follows. In Section \ref{sec:sysmod}, we present the GFDM system model and the new matrix characterization. We also derive some properties of GFDM matrices and present low-complexity transmitter implementations.
In Section \ref{sec:receiver}, we propose low-complexity ZF and MMSE receiver implementations.
In Section \ref{sec:complexity}, we present a thorough complexity analysis for GFDM implementations.
In Section \ref{sec:mmse}, optimal prototype transmit filters in terms of minimizing receiver MSEs are derived, and specific examples are provided.
In Section \ref{sec:oob}, we derive the analytical expression of power spectral density (PSD) and define the OOB leakage as a performance measure for the OOB radiation.
Simulation results are shown in Section \ref{sec:sim}. Finally, the study conclusion is provided in Section \ref{sec:conclusion}.

%
%
\tikzstyle{rect}=[draw, text centered, minimum height=2.8em]
\def\blockdist{1.3}
\begin{figure}[!t]
\centering
\begin{tikzpicture}[font=\fontsize{8}{8}\selectfont]
    \node (mod) [rect] {$\mA$};
    \path (mod)+(-0.6*\blockdist,0) node(modp){};
    \draw [->] (modp.center) -- node [above]{$\vd[l]$} (mod.west);
    \path (mod)+(0,0.7) node() {modulator};
    \path (mod)+(0.8*\blockdist,0) node(ps) [rect, text width=0.5em]{P\\/\\S};
    \draw [->] (mod) -- node [above]{$\vx[l]$} (ps.west);
    \path (ps)+(0.7*\blockdist,0) node(cp) [rect]{CP};
    \draw [->] (ps) -- node [above]{} (cp.west);
    \path (cp)+(\blockdist,0) node(channel) [rect]{$c[n]$};
    \draw [->] (cp) -- node [above]{$x[n]$} (channel.west);
    \path (channel)+(0,0.7) node() {channel};
    \path (channel)+(0.7*\blockdist,0) node(sum) [draw, circle, inner sep=0] {+};
    \draw [->] (channel) -- node [above]{} (sum.west);
    \path (sum.north)+(0,0.4*\blockdist)node (q) {$q[n]$};
    \draw [->] (q) -- (sum.north);
    \path (sum)+(0.8*\blockdist,0) node(rmcp) [rect, text width=1em]{r/m\\CP};
    \draw [->] (sum) -- node [above]{$y[n]$} (rmcp.west);
    \path (rmcp)+(0.7*\blockdist,0) node (sp) [rect, text width=0.5em]{S\\/\\P};
    \draw [->] (rmcp) -- node [above]{} (sp.west);
    \path (sp)+(0.8*\blockdist,0) node (dem) [rect]{$\mB$};
    \draw [->] (sp) -- node [above]{$\vy[l]$} (dem.west);
    \path (dem)+(0,0.7) node() {demodulator};
    \path (dem)+(0.7*\blockdist,0) node(demn){};
    \draw [->] (dem) -- node [above]{$\hat{\vd}[l]$} (demn);
\end{tikzpicture}
\caption{Block diagram of the transceiver. ("r/m" stands for "remove".)}
\label{fig:block}
\end{figure}
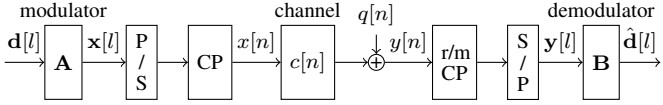

\textit{Notations:} Boldfaced capital letters denote matrices{,} and boldfaced lowercase letters are reserved for column vectors. We use $\modd{\cdot}_{D}$, $(\cdot)^*$, $(\cdot)^T$, and $(\cdot)^H$ to denote modulo ${D}$, complex conjugate, transpose, and Hermitian transpose, respectively. We also use $(\cdot)^{-H}$ to denote $((\cdot)^{-1})^H$. Given a matrix $\mA$, we denote by $[\mA]_{m,n}$, $[\mA]_{:,r}$, $\fnorm{\mA}$, $\vect(\mA)$, and $\mA^{\circ-1}$ its ($m,n$)th entry (zero-based indexing), $r$th column, Frobenius norm, column-wise vectorization, and Hadamard inverse (defined by $[\mA^{\circ-1}]_{m,n} = [\mA]_{m,n}^{-1}$, $\forall\; m,n$), respectively.
For any diagonal matrix $\mA$, $[\mA]_n$ denotes $[\mA]_{n,n}$.
For any matrices $\mA$ and $\mB$, $\mA\otimes\mB$ denotes their Kronecker product, and $\mA\circ\mB$ their Hadamard product. Given a vector $\vu$, we use $[\vu]_n$ to denote the $n$th component of $\vu$, $\|\vu\|$ the L2-norm of $\vu$, $\diag(\vu)$ the diagonal matrix containing $\vu$ on its diagonal, and $\boldsymbol{\Psi}(\vu)$ the circulant matrix whose first column is $\vu$.
Given square matrices $\mA_m$, $\forall\; 0\leq m<p$ for any positive integer $p$, we use $\blkdiag(\{\mA_m\}_{m=0}^{p-1})$ to denote a block diagonal matrix whose $m$th diagonal block is $\mA_m$.
We define $\mI_p$ to be the $p\times p$ identity matrix, $\mathbf{1}_p$ the $p\times 1$ vector of ones, $\mW_p$ the normalized $p$-point DFT matrix with $[\mW_p]_{m,n}=e^{-j2\pi mn/p}/\sqrt{p}$ for any positive integer $p$, and $\delta_{kl}$ the Kronecker delta.
We use $\angle C$ to denote the phase $\phi\in(-\pi,\pi]$ of a nonzero complex number $C$, and $\angle \mA$ the matrix such that $[\angle\mA]_{m,n}=\angle[\mA]_{m,n}$ for each entry.
For any set $\cA$, we denote its cardinality by $|\cA|$.
Finally, we use $\e\{\cdot\}$ to denote the expectation operator.


\section{Characterization of GFDM Systems}
\label{sec:sysmod}
GFDM is a block-based communication scheme as shown in Fig. \ref{fig:block} \cite{michailow14}. In a GFDM block, $M$ complex-valued subsymbols are transmitted on each of the $K$ subcarriers, so a total of $D=KM$ data symbols are transmitted. The data symbol vector $\vd[l]$ is decomposed as $\vd[l]=[d_{0,0}[l] \cdots d_{K-1,0}[l] \: d_{0,1}[l] \cdots d_{K-1,1}[l] \cdots d_{K-1,M-1}[l]]^T$, where $d_{k,m}[l]$ is the data symbol on the $k$th subcarrier and $m$th subsymbol in the $l$th block, taken from a complex constellation. Assume the data symbols are zero-mean and independent and identically distributed (i.i.d.) with symbol energy $E_S$, i.e., $\e\{\vd[l]\vd^H[n]\}=E_S \mI_D\delta_{ln}$. Each data symbol $d_{k,m}[l]$ is pulse-shaped by the vector $\vg_{k,m}$ whose $n$th entry is
\begin{equation}
    [\vg_{k,m}]_n = [\vg]_{\modd{n-mK}_D}e^{j2\pi kn/K}, n=0,1,\dots,D-1,\label{eq:gkmdef}
\end{equation}
where $\vg$ is a $D\times 1$ vector, referred to as the \emph{prototype transmit filter} \cite{michailow14}.
Let $\vx[l]=[x_0[l] \: x_1[l] \cdots x_{D-1}[l]]^T$ be the vector containing the transmit samples. Then, the GFDM modulator can be formulated as the transmitter matrix \cite{michailow14}
\begin{IEEEeqnarray}{rCl}
\mA=\left[\vg_{0,0} \cdots \vg_{K-1,0} \: \vg_{0,1} \cdots \vg_{K-1,1} \cdots \vg_{K-1,M-1}\right]\IEEEeqnarraynumspace \label{eq:gfdmmtx}
\end{IEEEeqnarray}
such that $\vx[l]=\mA\vd[l]$. 
The matrix $\mA$ as defined in (\ref{eq:gfdmmtx}) is called hereafter a \emph{GFDM matrix} with a \emph{prototype filter} $\vg$.
The vector $\vx[l]$ is further added a cyclic prefix (CP) before sending to the receiver through a linear time-invariant (LTI) channel. Details on the channel effects and the receiver are elaborated in Section \ref{sec:receiver}.

\subsection{Characterization of GFDM Matrices: Basic Definitions}

In the literature, GFDM transmitter matrices are often characterized by the prototype transmit filter $\vg$. Alternatively, in \cite{michailow12b, matthe16b, han16arxiv}, GFDM matrices have been parametrized by the frequency-domain prototype transmit filter $\vg_f=\sqrt{D}\mW_D\vg$, i.e., the $D$-point DFT of $\vg$.

In this paper, we propose an alternative means for characterizing a GFDM transmitter matrix, namely, the \emph{characteristic matrix} $\mG$ of size $K\times M$. 
We show that the proposed characterization is useful for understanding some important properties of GFDM transmitter matrices not easily derived in terms of the characterization of traditional time-domain or frequency-domain prototype filters. 
The proposed characterization is essentially equivalent to the discrete Zak transform (DZT) \cite{matthe14b,bolcskei97}, but all derivations in the paper do not require knowledge of the DZT.
A formal definition of this characterization of a GFDM transmitter matrix 
is given as follows.

\vspace{0.5em}

\begin{definition}[Characteristic matrix] \label{dfn:gfdmmtx}
Consider a $KM\times KM$ GFDM matrix $\mA$ in (\ref{eq:gfdmmtx}) with a prototype filter $\vg$.  
We define the \emph{characteristic matrix} $\mG$ of the GFDM matrix $\mA$ as
\begin{equation}
    \mG=\sqrt{D} \reshape(\vg, K, M) \mW_M,\label{eq:chrmtxdef}
\end{equation}
where $\reshape(\vg, K, M)$ is a $K\times M$ matrix whose $(k,m)$-entry is $[\vg]_{k+mK}$, $\forall\; 0\leq k<K$, $0\leq m<M$.
Moreover, the \emph{phase-shifted characteristic matrix} $\bar{\mG}$ of the GFDM matrix $\mA$ is defined as the $K\times M$ matrix whose $(k,m)$-entry is 
\begin{equation}
    [\bar{\mG}]_{k,m} = [\mG]_{k,m} e^{-j2\pi km/D}.\label{eq:pscm}
\end{equation}
\end{definition}
Finally, the \emph{energy} $\xi_G$ of the GFDM matrix $\mA$ is defined by $\xi_G=\fnorm{\mG}^2/D$.

\vspace{0.5em}

The following lemma would be useful for derivations of low-complexity transceiver implementations and optimal prototype filters later.

\vspace{0.5em}

\begin{lemma} \label{lmm:gfdmpoly}
Let $\mA$ be a GFDM matrix with a $K\times M$ characteristic matrix $\mG$, a $K\times M$ phase-shifted characteristic matrix $\bar\mG$, a $D\times 1$ prototype filter $\vg$, and energy $\xi_G$, where $D=KM$. Then, \\
(a) The prototype filter $\vg$ can be expressed as
$\vg=\vect\left(\mG \mW_M^H\right)/\sqrt{D}$.\\
(b) The frequency-domain prototype filter $\vg_f \triangleq \sqrt{D}\mW_D\vg$ can be expressed as $\vg_f = \vect( \bar{\mG}^T \mW_K)$.\\
(c) The matrix $\mA$ satisfies
\begin{IEEEeqnarray}{rCl}
    \mA &=& (\mW_M^H \otimes \mI_K) \diag(\vect(\mG)) (\mW_M \otimes \mW_K^H).\label{eq:cmimpl1}
\end{IEEEeqnarray}
(d) The energy $\xi_G$ satisfies $\xi_G=\|\vg\|^2$.
\end{lemma}
\begin{IEEEproof}
(a) The statement follows from the inverse operation of (\ref{eq:chrmtxdef}).\\
(b) According to (a), the prototype filter $\vg$ satisfies $[\vg]_{mK+k}=[\mG\mW_M^H]_{k,m}/\sqrt{D}$. Thus, $\vg_f$ satisfies
\begin{IEEEeqnarray}{rCl}
    [\vg_f]_{k'M+m'}&=&\sum_{k=0}^{K-1}\sum_{m=0}^{M-1}[\vg]_{mK+k}e^{-j2\pi (mK+k)(k'M+m')/D}\IEEEnonumber \\
    &=&\frac{1}{\sqrt{K}}\sum_{k=0}^{K-1}[\mG]_{k,m'}e^{-j2\pi k(k'M+m')/D}, \label{eq:hl}
\end{IEEEeqnarray}
$\forall\; 0\leq k'<K$, $0\leq m'<M$, i.e., $\vg_f = \vect( \bar{\mG}^T \mW_K)$.\\
(c) Using the famous matrix identity $\vect(\mA\mB\mC)=(\mC^T\otimes\mA)\vect(\mB)$\cite{laub04}, we first obtain that $(\mW_M^H \otimes \mI_K)\vect(\mG) = \vect(\mI_K\mG(\mW_M^H)^T) = \vect(\mG\mW_M^H) = \sqrt{D}\vg.$
Then, the zeroth column of the right-hand side of (\ref{eq:cmimpl1}) is $(\mW_M^H \otimes \mI_K)\diag(\vect(\mG))\cdot\frac{1}{\sqrt{D}}{\bf 1}_D=(\mW_M^H \otimes \mI_K)\vect(\mG)/\sqrt{D}=\vg$, i.e., the prototype filter of $\mA$. The equality of the other columns of both sides in (\ref{eq:cmimpl1}) can be verified similarly, by noting that the $(k+mK)$th column of $\mW_M\otimes \mW_K^H$ is $[\mW_M]_{:,m}\otimes[\mW_K]_{:,k}.$\\
\ignore{
Since $\diag([(\mW_M \otimes \mW_K^H)]_{:,k+mK}) = \mathbf{\Phi}_{-m}^{(M)} \otimes \mathbf{\Phi}_k^{(K)}$, we have
$[\diag(\vect(\mG)) (\mW_M \otimes \mW_K^H)]_{:,k+mK} = \vect( \mathbf{\Phi}_k^{(K)} \mG \mathbf{\Phi}_{-m}^{(M)}) / \sqrt{D}$.
Besides, $(\mW_M^H \otimes \mI_K) \vect(\mM) = \vect\left( \mM \mW_M^H \right)$ for any $K\times M$ matrix $\mM$ and $\mathbf{\Phi}_{-m}^{(M)} \mW_M^H = \mW_M^H \mP_m$. Thus, we derive that the $(k+mK)$th column of the right hand side of (\ref{eq:cmimpl1}) is $\vect( \mathbf{\Phi}_k^{(K)} \mG \mW_M^H \mP_m ) / \sqrt{D}$.}
(d) The proof is trivial in view of Lemma \ref{lmm:gfdmpoly}(a) and Parseval's theorem.
\end{IEEEproof}

\vspace{0.5em}

Lemmas \ref{lmm:gfdmpoly}(a) and \ref{lmm:gfdmpoly}(b) indicate the one-to-one correspondence among $\mG$, $\vg$, and $\vg_f$ and are useful in developments later in this paper.
It is noted that a mathematically equivalent form of them can also be derived from the definition and frequency-domain expression of the DZT \cite{bolcskei97}.
The statements and proofs provided here, however, do not require knowledge of the DZT. 
Lemma \ref{lmm:gfdmpoly}(c) is a simplified form of the decomposition proposed in \cite{lin16}, and we give a simple alternative proof above. We will use (\ref{eq:cmimpl1}) to develop transceiver implementations.
Finally, Lemma \ref{lmm:gfdmpoly}(d) shows that the energy of $\mA$ is simply the energy of the prototype filter $\vg$, which can also be proved by unitarity of the DZT \cite{bolcskei97}.










\subsection{GFDM Transmitter Implementations}

\ignore
{
Now we present the detailed models of the transmitter, channel, and receiver. Among them, three types of implementation of the transmitter and receiver will be shown.
}

As presented earlier in this paper, the transmitter simply modulates the data symbol vector by
\begin{equation}
    \vx[l]=\mA\vd[l].\label{eq:xad}
\end{equation}
Then, $\vx[l]$ is passed through a parallel-to-serial (P/S) converter, and a CP of length $L$ is added, as shown in Fig. \ref{fig:block}. Denote $\cK\subseteq\{0, 1, \dots, K-1\}$ and $\cM\subseteq\{0, 1, \dots, M-1\}$ as the set of subcarrier indices and set of subsymbol indices, respectively, that are actually used. The digital baseband transmit signal of GFDM can then be expressed as
\begin{IEEEeqnarray}{rCl}
    x[n] &=& \sum_{l=-\infty}^{\infty}\sum_{k\in\cK}\sum_{m\in\cM} d_{k,m}[l] g_m[n-lD'] e^{j2\pi \frac{k(n-lD')}{K}},\IEEEeqnarraynumspace\label{eq:dbts}
\end{IEEEeqnarray}
where $D'=D+L$ and
\begin{equation}
    g_m[n] = \left\{ \begin{array}{cl}
        [\vg]_{\modd{n-mK-L}_D}, & n = 0, 1, \dots, D'-1 \\
        0, & \text{otherwise}
    \end{array}\right..\label{eq:gmn}
\end{equation}
In most instances of this paper, we omit the block index "$[l]$" for notational brevity.

For the implementation of the transmitter matrix $\mA$, two types pertaining to the conventional time \cite{michailow14} and frequency \cite{matthe16b} domains, respectively, are found in the literature.
In this paper, we propose another implementation based on the characteristic matrix.
These implementations are described as follows:

\subsubsection{Direct implementation}
The matrix multiplication in (\ref{eq:xad}) is directly implemented, which can be considered a time-domain implementation that deals with the prototype filter $\vg$ directly \cite{michailow14}.

\subsubsection{Frequency-domain implementation}
Previous frequency-domain implementations \cite{michailow12b, matthe16b} have been proposed for complexity reduction. The transmit signal is produced with
\begin{equation}
    \vx = \frac{1}{\sqrt{K}} \mW_D^H \sum_{k\in\cK} \mP^{(k)} \diag(\vg_f) \mR \mW_M \vd_k,
\end{equation}
where $\vd_k=[d_{k,0} \: d_{k,1} \cdots d_{k,M-1}]^T$, $\mR=\mathbf{1}_K \otimes \mI_M$, and $\mP^{(k)} = \boldsymbol{\Psi}(\vp^{(k)}) \otimes \mI_M$,
with $\vp^{(k)}$ being the $K\times 1$ vector equal to the $k$th column of $\mI_K$.

\subsubsection{Characteristic-matrix-domain implementation}
\begin{figure}[!t]
\centering

\includegraphics[width=3.4in,height=1.5in]{\figpath/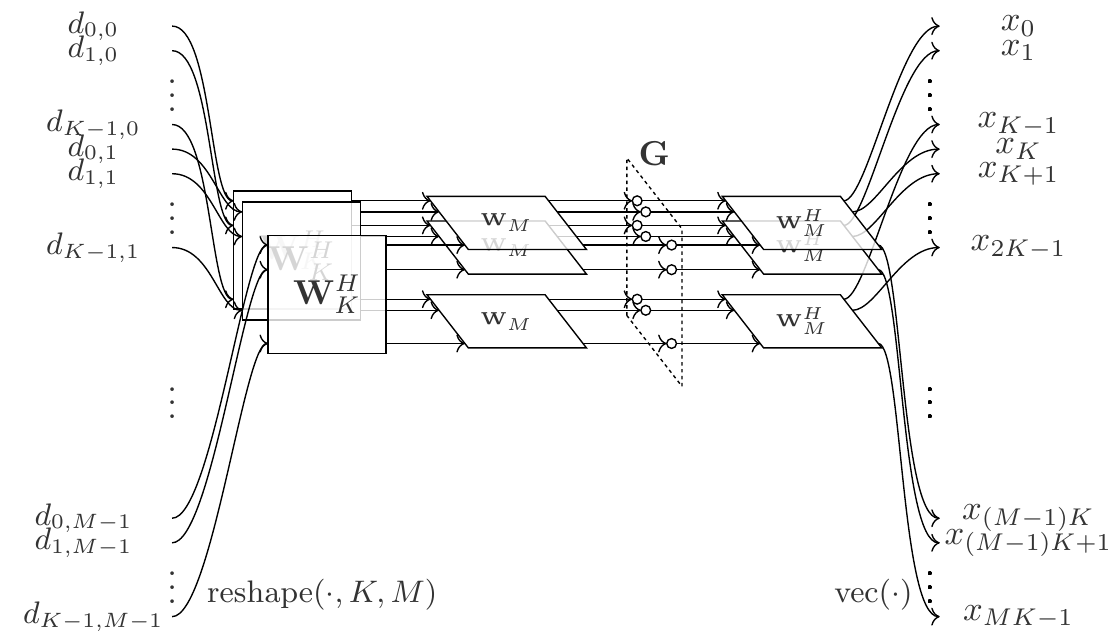}

\caption{Characteristic-matrix-domain Form-1 transmitter implementation.}
\label{fig:CMTx1}

\end{figure}
\begin{figure}[!t]
\centering

\includegraphics[width=3.4in, height=1.5in]{\figpath/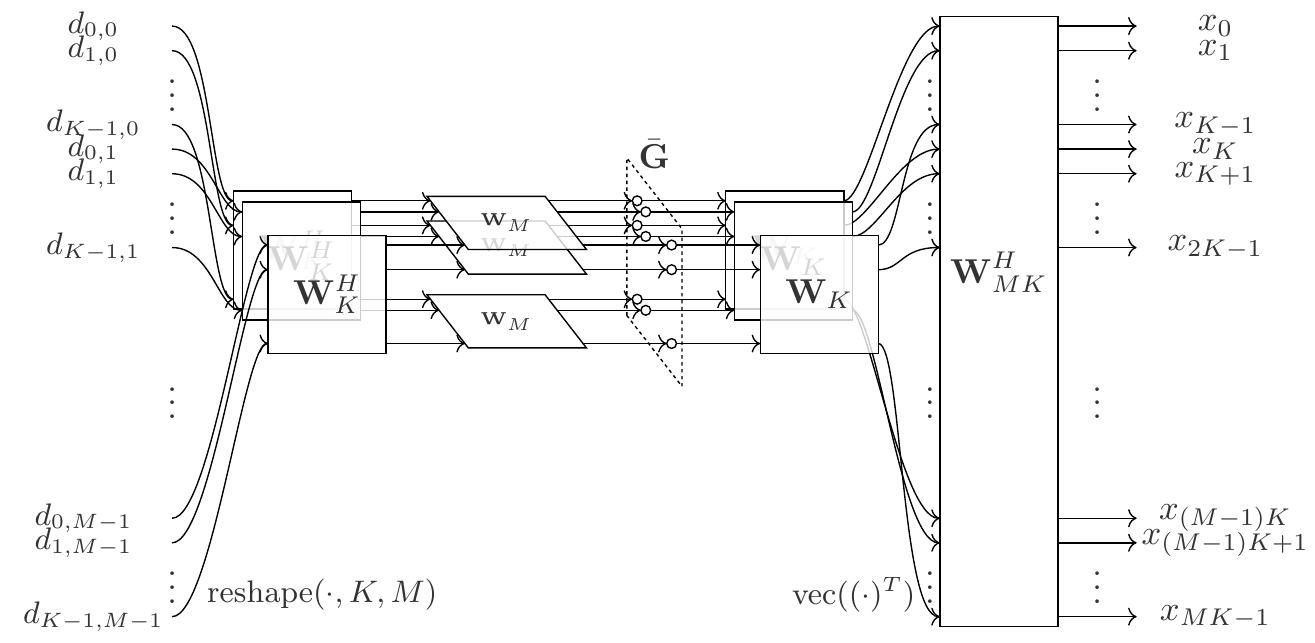}

\caption{Characteristic-matrix-domain Form-2 transmitter implementation.}
\label{fig:CMTx2}
\end{figure}

We propose two forms of \emph{characteristic-matrix-domain implementation}.
Using Lemma \ref{lmm:gfdmpoly}(c), we obtain a transmitter implementation based on (\ref{eq:cmimpl1}), which we call Form-1 implementation, as shown in Fig. \ref{fig:CMTx1}.
An alternative form of decomposition of the transmitter matrix that exploits the phase-shifted characteristic matrix $\bar{\mG}$ is formulated as
\begin{IEEEeqnarray}{rCl}
    \mA &=& \mW_D^H \mathbf{\Pi} (\mI_M \otimes \mW_K) \diag(\vect(\bar{\mG})) (\mW_M \otimes \mW_K^H),\IEEEeqnarraynumspace\label{eq:cmimpl2}
\end{IEEEeqnarray}
where $\mathbf{\Pi}$ is the $D\times D$ permutation matrix defined by
\begin{IEEEeqnarray}{rCl}
    [\mathbf{\Pi}]_{kM+m,nK+l} &=& \delta_{kl} \delta_{mn},\IEEEeqnarraynumspace\label{eq:pidef}
\end{IEEEeqnarray}
$\forall\; 0\leq k,l<K$, $0\leq m,n<M$. The matrix $\mathbf{\Pi}$ can be understood through the identity $\vect(\mM^T) = \mathbf{\Pi} \vect(\mM)$, where $\mM$ is any $K\times M$ matrix. We obtain (\ref{eq:cmimpl2}) by using (\ref{eq:cmimpl1}) and the fact that a $KM$-point DFT can be decomposed into a $K$-point DFT, an $M$-point DFT, and twiddle factors of the form $e^{-j2\pi km/D}$, which are incorporated into $\bar{\mG}$. Eq. (\ref{eq:cmimpl2}) corresponds to the implementation shown in Fig. \ref{fig:CMTx2}, which we call Form-2 implementation. 
The complexity of both forms are in $O(KM\log KM)$.
Yet, as will be seen in Section \ref{sec:complexity}, the complexity of Form-1 transmitter is slightly lower than that of Form-2 transmitter, while the Form-2 structure based on the decomposition in (\ref{eq:cmimpl2}) is advantageous for receiver implementation.

\subsection{Unitary and Invertible GFDM Matrices}

With the characteristic-matrix-domain implementation, we can also easily identify the class of unitary GFDM matrices as follows.

\vspace{0.5em}

\begin{theorem}[Unitary GFDM matrices] \label{thm:unitary}
Let $\mA$ be a GFDM matrix with a $K\times M$ characteristic matrix $\mG$. Then, $\mA$ is unitary if and only if $\mG$ contains unit-magnitude entries: $|[\mG]_{k,l}|=1\;\forall\; 0\leq k< K, 0\leq l< M$. An equivalent condition is that its phase-shifted characteristic matrix $\bar{\mG}$, as defined in (\ref{eq:pscm}), contains unit-magnitude entries: $|[\bar{\mG}]_{k,l}|=1\;\forall\; 0\leq k< K, 0\leq l< M$.
\end{theorem}
\begin{IEEEproof}
Since $\mW_M^H \otimes \mI_K$ and $\mW_M \otimes \mW_K^H$ in (\ref{eq:cmimpl1}) are both unitary, $\mA$ is unitary if and only if the diagonal matrix $\diag(\vect(\mG))$ is unitary. This is the case if and only if $|[\mG]_{k,l}|=1\;\forall\; 0\leq k< K, 0\leq l< M$. Finally, we have the equivalent condition since $|[\mG]_{k,l}|=|[\bar{\mG}]_{k,l}|$, $\forall\; k,l$.
\end{IEEEproof}

\vspace{0.5em}

Observing the result in Theorem \ref{thm:unitary}, we call a prototype filter $\vg$ a \emph{constant-magnitude-characteristic-matrix (CMCM)} filter if the corresponding characteristic matrix contains constant-magnitude entries, i.e., corresponding to a scalar multiple of a unitary GFDM matrix. We will show that CMCM filters are solutions to several of our problems in minimizing the receiver MSE, and are an important class of filters for GFDM.

The following theorem expresses the conditions for the non-singularity of a GFDM matrix in terms of its characteristic matrix and related properties. Later in this paper, the theorem is shown to be very useful in our study on a GFDM receiver.

\vspace{0.5em}

\begin{theorem}[Properties of $\mA^{-1}$] \label{thm:prop}
Let $\mA$ be a GFDM matrix with a $K\times M$ characteristic matrix $\mG$. Then, \\
(a) $\mA$ is invertible if and only if $\mG$ has no zero entries. \\
(b) If $\mA$ is invertible, then $\mA^{-H}$ is also a GFDM matrix whose characteristic matrix $\mH$ satisfies  $[\mH]_{k,l}=1/[\mG]^*_{k,l},\forall k,l$, i.e.,

\begin{equation}
    \mH=(\mG^*)^{\circ-1}. \label{eq:ghpoly}
\end{equation}
(c) If $\mA$ is invertible, the squared norm of each row of $\mA^{-1}$ equals the energy of $\mA^{-H}$, $\xi_H=\fnorm{\mH}^2/D$.
\end{theorem}
\begin{IEEEproof}
(a) According to (\ref{eq:cmimpl1}), $\mA$ is invertible if and only if $\mG$ has no zero entries since $\mW_M^H \otimes \mI_K$ and $\mW_M \otimes \mW_K^H$ are both unitary matrices.\\
(b) According to (\ref{eq:cmimpl1}), if $\mA$ is invertible,
\begin{IEEEeqnarray}{rCl}
    \mA^{-H} &=& (\mW_M^H \otimes \mI_K) (\diag(\vect(\mG))^{-H} (\mW_M \otimes \mW_K^H).\IEEEeqnarraynumspace
\end{IEEEeqnarray}
In other words, $\mA^{-H}$ is a GFDM matrix whose characteristic matrix $\mH$ satisfies (\ref{eq:ghpoly}).\\
(c) According to (b), $\mA^{-H}$ is a GFDM matrix. Since the norm of each column of a GFDM matrix equals the norm of its prototype filter, the result follows from Lemma \ref{lmm:gfdmpoly}(d).
\end{IEEEproof}

\vspace{0.5em}

The condition for the singularity of $\mA$ is also found in \cite{lin16}. In \cite{matthe14b}, Gabor analysis results \cite{zibulski93, bolcskei97} and DZT \cite{bolcskei97} were applied to obtain a similar statement in Theorem \ref{thm:prop}(b).
Our derivations, however, involve only basic linear algebra and DFT, making the properties more accessible to general readers.

\section{GFDM Receiver Implementations}\label{sec:receiver}
In this section, we complete our description of the GFDM system model as illustrated in Fig. \ref{fig:block}, and propose a new form of low-complexity implementation of ZF and MMSE receivers: the characteristic-matrix-domain implementation.

As shown in Fig. \ref{fig:block}, the received signal after transmission through a wireless channel is modeled as an LTI system $y[n]=c[n]*x[n]+q[n]$, where $c[n]$ is the channel impulse response, and $q[n]$ is the complex AWGN with variance $N_0$. When $c[n]=\delta_{n0}$, the channel reduces to an AWGN channel. More generally, we consider a multipath channel with arbitrary coefficients $c[n]$. The channel order is assumed not to exceed the CP length; that is, $c[n]=0$ for all $n$ such that $n<0$ or $n>L$.
The received samples after CP removal and serial-to-parallel (S/P) conversion are collected as $\vy[l]=[y_0[l] y_1[l] \cdots y_{D-1}[l]]^T$.
The transfer function from the transmitted block $\vx[l]$ to the received block $\vy[l]$ is
\begin{equation} 
\vy[l] = \mC \vx[l] + \vq[l], \label{eq:channel}
\end{equation}
where $\mC$, the channel circular convolution matrix \cite{michailow14}, equals the circulant matrix $\boldsymbol{\Psi}([c[0] c[1] \cdots c[D-1]]^T)$ \cite{lin10}.
As there is no inter-block interference, the index ``$[l]$'' is omitted in most parts of the following developments.
Since a circulant matrix can be diagonalized by the DFT matrix, we have 
\begin{equation}
\mC = \mW_D^H \mD_C \mW_D, \label{eq:circdiag}
\end{equation}
where $\mD_C = \diag([C_0 C_1 \cdots C_{D-1}]^T)$ with $C_l=\sum_{n=0}^{D-1} c[n] e^{-j2\pi nl/D}$ being the $D$-point DFT of $c[n]$.

\ignore
{For (static) multipath channels, $c[n]$ is deterministic. For Rayleigh fading channels, $c[n]$ are independent circularly symmetric complex Gaussian with variance $N^c_n$ for $n=0,1,\dots,L$. 
If the channel impulse response is of order $L_c<L$, it corresponds to our model with $N^c_n=0, \forall\; L_c<n\leq L$. Let the energy of $c[n]$ be normalized to unity.}

From (\ref{eq:xad}) and (\ref{eq:channel}), we can express the received block in terms of the source data symbol vector as
\begin{equation}
    \vy = \mC\mA\vd + \vq. \label{eq:yCAd}
\end{equation}
The receiver is responsible for obtaining the estimated data symbol vector $\hat{\vd}$ given the received block $\vy$.
In the literature, several standard types of receivers have been discussed \cite{michailow14, matthe14b}, including MF, ZF, and linear MMSE receivers. Note that when unitary GFDM transmitter matrices are used, an MF receiver is equivalent to a ZF receiver because $\mA^{-1}=\mA^H$ if $\mA$ is unitary. We study ZF and MMSE receivers in this paper.

\begin{figure}[!t]
\centering

\includegraphics[width=3.4in, height=1.5in]{\figpath/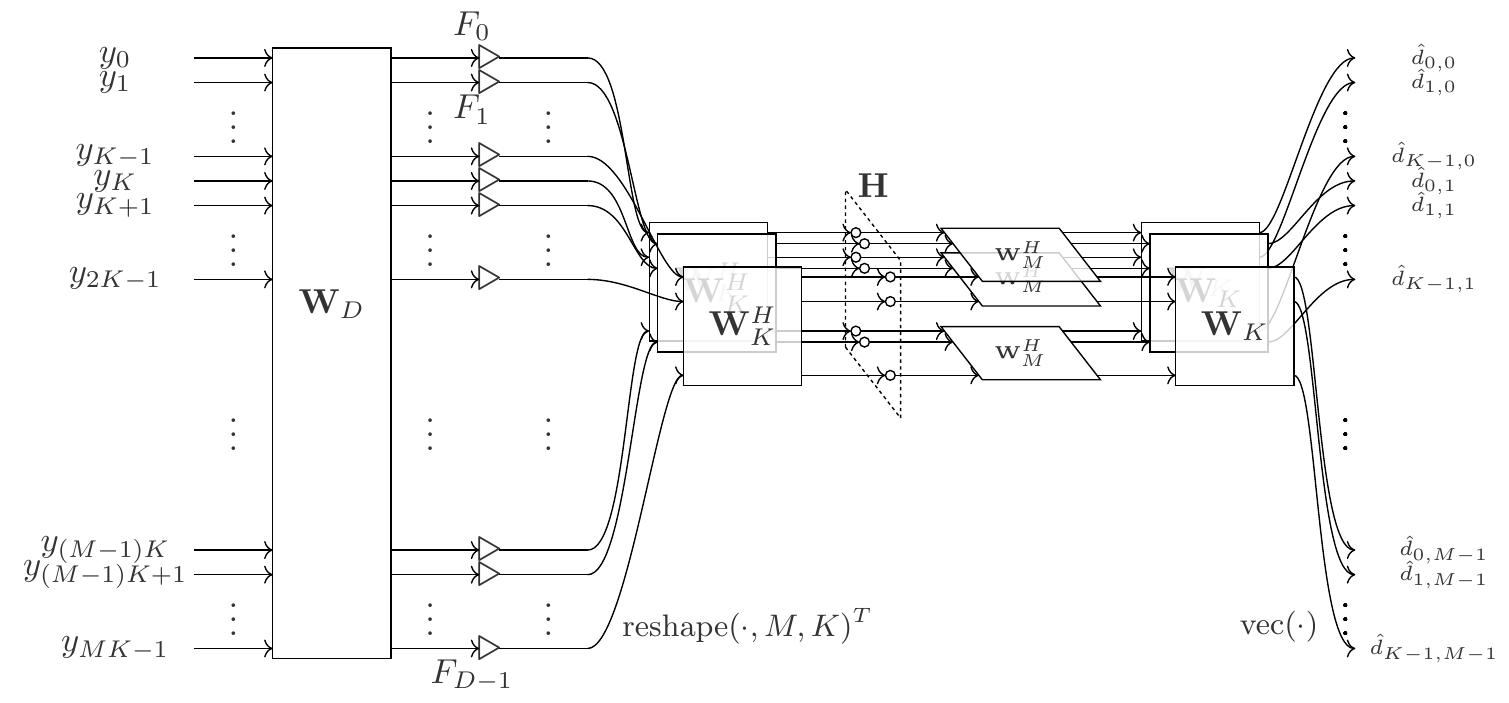}

\caption{Characteristic-matrix-domain Form-2 receiver implementation.}
\label{fig:CMRx2}
\end{figure}

\subsection{Low-Complexity ZF Receivers}
In the ZF receiver, the demodulator $\mB_{\mathrm{ZF}}$ is formulated as a GFDM receiver matrix $\mA^{-1}$ multiplied by an equalizer $\mC^{-1}$. The estimated data symbol vector is
\begin{equation}
    \hat{\vd}=\mB_{\mathrm{ZF}}\vy=\mA^{-1}\mC^{-1}\vy=\vd+\mA^{-1}\mC^{-1}\vq. \label{eq:estsym}
\end{equation}
Note that the ZF receiver exists when both $\mA$ and $\mC$ are invertible. 
Theorem \ref{thm:prop}(b) implies that $\mA^{-1}$ is just a Hermitian transpose of another GFDM matrix. Combined with the fact that $\mC$ is diagonalizable by $\mW_D$, low-complexity implementations for the ZF receiver based on the forms in (\ref{eq:cmimpl1}) and (\ref{eq:cmimpl2}) are readily available.
Particularly, we obtain the ZF receiver Form-1 implementation
\begin{IEEEeqnarray}{rCl}
    \mB_{\mathrm{ZF}} &=& (\mW_M^H \otimes \mW_K) \mD_G^{-1} (\mW_M \otimes \mI_K) \mW_D^H \mD_C^{-1} \mW_D, \IEEEeqnarraynumspace\label{eq:CMRx1}
\end{IEEEeqnarray}
where $\mD_G=\diag(\vect(\mG))$, and the ZF receiver Form-2 implementation
\begin{IEEEeqnarray}{rCl}
    \mB_{\mathrm{ZF}} &=& (\mW_M^H \otimes \mW_K) \bar{\mD}_G^{-1} (\mI_M \otimes \mW_K^H) \mathbf{\Pi}^T \mD_C^{-1} \mW_D, \IEEEeqnarraynumspace\label{eq:CMRx2}
\end{IEEEeqnarray}
where $\bar{\mD}_G\diag(\vect(\bar{\mG}))$. 
The block diagram of a Form-2 receiver is shown in Fig. \ref{fig:CMRx2}, with $F_l = 1/C_l$, $\forall\; 0\leq l<D$ and $\mH = \bar{\mG}^{\circ-1}$ therein.
Although the complexity of both forms is in $O(KM\log KM)$, we show in Section \ref{sec:complexity} that Form 2 is generally of lower complexity. \ignore{This is because Form 2 requires only one $D$-point DFT, but Form 1 requires two $D$-point DFTs. }
Yet, under the special case of the AWGN channel, Form-1 implementation is advantageous since it is simplified to
\begin{IEEEeqnarray}{rCl}
    \mA^{-1} &=& (\mW_M^H \otimes \mW_K) \mD_G^{-1} (\mW_M \otimes \mI_K),\IEEEeqnarraynumspace
\end{IEEEeqnarray}
which does the reverse operation of Fig. \ref{fig:CMTx1}.

The frequency-domain implementation can be used for ZF receivers.
It is proposed that estimated data symbols for the $k$th subcarrier in the ZF receiver are given by \cite{michailow12b, matthe16b}{, \cite{wei16}}
\begin{equation}
    \hat{\vd}_k = \frac{1}{\sqrt{K}} \mW_M^H \mR^T \diag(\vh_f) (\mP^{(k)})^T \mD_C^{-1} \mW_D \vy,\label{eq:FDRx}
\end{equation}
where $\vh_f$ is the ZF frequency-domain prototype receive filter. \ignore{(The use of prototype receive filter hints that $\mA^{-H}$ is also a GFDM matrix, which will be shown later.)}

\subsection{Low-Complexity MMSE Receivers}

For an MMSE receiver, the existence of a low-complexity implementation at the order $O(KM\log KM)$ has not been well studied previously except in the case of an AWGN channel \cite{matthe16b, farhang16}.
 Assuming $\e\{\vd\vd^H\}=E_S \mI_D$ (i.e., all subcarriers are subsymbols are allocated with data)\footnote{If this is not the case, then the subsequent derivations on the MMSE receiver are not exact and may need modification in the future.},
the MMSE receiver for (\ref{eq:yCAd}) can be modeled as \cite{lin10}
\begin{equation}
\mB_{\mathrm{MMSE}}=\mA^H\mC^H\left[\mC\mA\mA^H\mC^H + \gamma^{-1}\mI_D \right]^{-1}, \label{eq:mmmserxmat}
\end{equation}
where $\gamma=E_S/N_0$ is the signal-to-noise ratio (SNR), and
\begin{equation}
    \hat{\vd}=\mB_{\mathrm{MMSE}}\vy. \label{eq:estsymmmse}
\end{equation}
When both $\mA$ and $\mC$ are invertible, (\ref{eq:mmmserxmat}) reduces to  \cite{matthe14b}
\begin{equation}
\mB_{\mathrm{MMSE}}=\left[\mC\mA+\gamma^{-1}(\mC\mA)^{-H}\right]^{-1}. \label{eq:mmmserxmat2}
\end{equation}

Either (\ref{eq:mmmserxmat}) or (\ref{eq:mmmserxmat2}) involves the inversion of a matrix that is not a GFDM matrix, so Theorem \ref{thm:prop} does not apply here to the reduction of the implementation complexity. A direct implementation requires a complexity of $O(K^3M^3)$ and is often not a desirable solution.
Also, the frequency-domain implementation \cite{matthe16b} is not applicable to the MMSE receiver in general since (\ref{eq:mmmserxmat}) cannot be simplified to the form in (\ref{eq:FDRx}).  

We propose to use the structure depicted in Fig. \ref{fig:CMRx2} in our study of a potential MMSE receiver, where coefficients $F_k$ and entries of $\mH$ are to be designed. The following theorem provides the necessary and sufficient conditions on which an MMSE receiver can be implemented with such a form.

\vspace{0.5em}

\begin{theorem}\label{thm:mmselow}
Let $\mA$ be a nonsingular GFDM matrix with a $K\times M$ phase-shifted characteristic matrix $\bar{\mG}$, $\mC$ be a $D\times D$ nonsingular circulant matrix, $\gamma$ be a positive real number, $\mD_C=\mW_D\mC\mW_D^H$, and $C_l=[\mD_C]_{l}, \forall\; 0\leq l< D$, where $D=KM$. Then, there exist $D\times D$ nonsingular diagonal matrices $\mD_1, \mD_2$ such that $\mB_{\mathrm{MMSE}}$ defined in (\ref{eq:mmmserxmat2}) satisfies
\begin{IEEEeqnarray}{rCl}
    \mB_{\mathrm{MMSE}}=(\mW_M^H \otimes \mW_K) \mD_2^{-1} (\mI_M \otimes \mW_K^H) \mathbf{\Pi}^T \mD_1^{-1} \mW_D\IEEEeqnarraynumspace\label{eq:lowcomp}
\end{IEEEeqnarray}
if and only if $\forall\; 0\leq m< M$, either (a) $|[\bar{\mG}]_{k,m}|$ is a constant in $k$, or (b) $|C_{kM+m}|$ is a constant in $k$, or both, where $\mathbf{\Pi}$ is defined in (\ref{eq:pidef}).
\end{theorem}
\begin{IEEEproof}
Let $\mQ = \mI_M \otimes \mW_K$.
Note that for any $D\times D$ diagonal matrix $\mD=\diag(\vs)$, $\mD' \triangleq \mathbf{\Pi}^T \mD \mathbf{\Pi}$ is also a diagonal matrix with $\mD' = \diag(\mathbf{\Pi}^T\vs)$.
Using this property and (\ref{eq:cmimpl2})(\ref{eq:circdiag}), one can show that
\begin{IEEEeqnarray}{rCl}
    \left[\mC\mA+\gamma^{-1}(\mC\mA)^{-H}\right]^{-1} &=& (\mW_M^H \otimes \mW_K) \mE^{-1} \mathbf{\Pi}^T \mW_D\IEEEeqnarraynumspace\label{eq:bmmselow}
\end{IEEEeqnarray}
where $\mE$ is defined as
\begin{IEEEeqnarray}{rCl}
    \mE &=& \mD_C' \mQ \bar{\mD}_G +  \gamma^{-1} \mD_C'^{-H} \mQ \bar{\mD}_G^{-H},\IEEEeqnarraynumspace\label{eq:emtxdef}
\end{IEEEeqnarray}
$\bar{\mD}_G=\diag(\vect(\bar{\mG}))$, $\mD_C' = \mathbf{\Pi}^T \mD_C \mathbf{\Pi}$ is a diagonal matrix with $\mD_C' = \diag(\mathbf{\Pi}^T\vc_f)$, and $\vc_f=[C_0 C_1 \cdots C_{D-1}]^T$. Noting that $\mW_M^H \otimes \mW_K$, $\mathbf{\Pi}^T$, and $\mW_D$ in (\ref{eq:bmmselow}) are all unitary, and that $\mathbf{\Pi}^T\mD_1^{-1} = \mD_3^{-1}\mathbf{\Pi}^T$ if we define $\mD_3$ as $\mD_3 = \mathbf{\Pi}^T \mD_1 \mathbf{\Pi}$, we determine that (\ref{eq:lowcomp}) is satisfied if and only if there exist nonsingular $D\times D$ diagonal matrices $\mD_3, \mD_2$ such that $\mE = \mD_3 \mQ \mD_2$.
Let $\vu_m, \tilde{\vu}_m, \vv_m, \tilde{\vv}_m, \vw_m, \vz_m$ be $K\times 1$ vectors $\forall\; 0\leq m< M$ such that
$\diag([\vu_0^T \cdots \vu_{M-1}^T]^T) = \mD_C'$, $\diag([\tilde{\vu}_0^T \cdots \tilde{\vu}_{M-1}^T]^T) = \mD_C'^{-H}$, $\diag([\vv_0^T \cdots \vv_{M-1}^T]^T) = \bar{\mD}_G$, $\diag([\tilde{\vv}_0^T \cdots \tilde{\vv}_{M-1}^T]^T) = \bar{\mD}_G^{-H}$,
\begin{equation}
    \diag([\vw_0^T \cdots \vw_{M-1}^T]^T) = \mD_3, \diag([\vz_0^T \cdots \vz_{M-1}^T]^T) = \mD_2.\IEEEeqnarraynumspace \label{eq:wzdef}
\end{equation}
Noting that $\mQ = \mI_M \otimes \mW_K = \blkdiag(\{\mW_K\}_{m=0}^{M-1})$, we obtain that $\mD_3 \mQ \mD_2 = \blkdiag(\{(\vw_m\vz_m^T) \circ \mW_K\}_{m=0}^{M-1})$ and $\mE = \blkdiag(\{\mF_m \circ \mW_K\}_{m=0}^{M-1})$, where $\forall\; 0\leq m< M$,
\begin{equation}
    \mF_m=[\vu_m\; \gamma^{-1}\tilde{\vu}_m] [\vv_m\; \tilde{\vv}_m]^T.\label{eq:fmtx}
\end{equation}
Since for both $\mE$ and $\mQ$, each block diagonal submatrix is a full matrix without any zero entry, $\mE = \mD_3 \mQ \mD_2$ is satisfied if and only if $\mF_m = \vw_m\vz_m^T$ is satisfied $\forall\; 0\leq m< M$.
For any given $m$, if condition (a) is satisfied:
$|[\vv_m]_k| = |[\bar{\mG}]_{k,m}|$ is a constant in $k$, then $\tilde{\vv}_m = |[\bar{\mG}]_{0,m}|^{-2} \vv_m$ and we can choose
\begin{equation}
    \vw_m = \vu_m + (\gamma |[\bar{\mG}]_{0,m}|^2)^{-1}\tilde{\vu}_m, ~~\vz_m = \vv_m,\label{eq:constantg}
\end{equation}
to make $\mF_m=\vw_m\vz_m^T$; if condition (b) is satisfied: 
$|[\vu_m]_k| = |C_{kM+m}|$ is a constant in $k$, then $\tilde{\vu}_m = |C_m|^{-2} \vu_m$ and we can choose
\begin{equation}
    \vw_m = \vu_m,~~ \vz_m = \vv_m + (\gamma |C_m|^2)^{-1}\tilde{\vv}_m.\label{eq:constantc}
\end{equation} to make $\mF_m = \vw_m\vz_m^T$. 
It is now clear that for any $m$, if at least one of (a) and (b) is satisfied, then there exist $\vw_m, \vz_m$, and consequently, $\mD_2,\mD_3$, such that $\mE$ in (\ref{eq:emtxdef}) satisfies $\mE=\mD_3\mQ\mD_2$.
Conversely, assume that $\mF_m = \vw_m\vz_m^T$ is satisfied $\forall\; 0\leq m< M$, but that both conditions (a) and (b) are not satisfied for some $m$, say, $m_0$. Then, both sets $\{\vu_{m_0}, \tilde{\vu}_{m_0}\}$ and $\{\vv_{m_0}, \tilde{\vv}_{m_0}\}$ are linearly independent.
Thus, $\rank(\mF_{m_0}) = 2$, which can be proved by, e.g., Sylvester’s law of nullity \cite{matsaglia74}.
This contradicts to the assumption $\mF_{m_0} = \vw_{m_0}\vz_{m_0}^T$.
\end{IEEEproof}

\vspace{0.5em}

Theorem \ref{thm:mmselow} implies that a unitary GFDM matrix and the AWGN channel are two sufficient (but not necessary) conditions for the existence of the low-complexity MMSE receiver implementation in the form of Fig. \ref{fig:CMRx2}.
Specifically, assuming $\mC$ in Theorem \ref{thm:mmselow} is the channel circulant matrix, we obtain that $|C_{kM+m}|$ is constant in $k$ for all $m$ under the AWGN channel. Thus, according to (\ref{eq:constantc}), the MMSE receiver under the AWGN channel can be implemented as shown in Fig. \ref{fig:CMRx2}, with $F_l = 1$, $\forall\; 0\leq l<D$ and
$\mH = (\bar{\mG}+\gamma^{-1}(\bar{\mG}^*)^{\circ-1})^{\circ-1}$
therein.
For the more practical case where $|C_{kM+m}|$ is non-constant in $k$ for all $m$, Theorem \ref{thm:mmselow} implies that a sufficient condition for a low-complexity MMSE receiver implementation in the form of Fig. \ref{fig:CMRx2} is that $|[\bar{\mG}]_{k,m}|$ is a constant in both $k$ and $m$, i.e., using a unitary GFDM matrix $\mA$ up to a scale factor, or equivalently, a CMCM filter, in view of Theorem \ref{thm:unitary}. In this case, each $|[\bar{\mG}]_{k,m}|^2$ equals the energy $\xi_G$ of $\mA$, and according to (\ref{eq:constantg}), we have the Form-2 implementation of the MMSE receiver shown in Fig. \ref{fig:CMRx2}, with $F_l = 1/(C_l+(\gamma \xi_G C_l^*)^{-1})$, $\forall\; 0\leq l<D$ and $\mH = \bar{\mG}^{\circ-1}$ therein.

\subsection{Low-Complexity Approximated MMSE Receivers}
\label{ssc:ammse}
If neither conditions (a) nor (b) in Theorem \ref{thm:mmselow} are satisfied for some $m$, then it is impossible to find $\mD_3, \mD_2$ such that $\mE = \mD_3 \mQ \mD_2$, where $\mE$ is defined in (\ref{eq:emtxdef}) and $\mQ = \mI_M \otimes \mW_K$.
In this case, an exact MMSE receiver cannot be implemented as shown in Fig. \ref{fig:CMRx2}, but we propose using an approximated MMSE receiver based on the same structure.
Specifically, we minimize the Frobenius norm $\fnorm{\mE - \mD_3 \mQ \mD_2}$ by using low-rank matrix approximations.
Since $\mE = \blkdiag(\{\mF_m \circ \mW_K\}_{m=0}^{M-1})$, $\mD_3 \mQ \mD_2 = \blkdiag(\{(\vw_m\vz_m^T) \circ \mW_K\}_{m=0}^{M-1})$, and $\mW_K$ contains constant-magnitude entries, an equivalent condition is minimizing $\fnorm{\mF_m - \vw_m\vz_m^T}, \forall\; 0\leq m< M$, where $\vw_m$, $\vz_m$, and $\mF_m$ are defined in (\ref{eq:wzdef}) and (\ref{eq:fmtx}). By performing the singular value decomposition (SVD) of $\mF_m$ for each $m$, we obtain
$\mF_m = \mU_m \mathbf{\Sigma}_m \mV_m^H$,
where $\mU_m$ and $\mV_m$ are $D\times D$ unitary matrices, and $\mathbf{\Sigma}_m = \diag([s_m^{(1)} s_m^{(2)} 0 \cdots 0]^T)$ with $s_m^{(1)}\geq s_m^{(2)}$.
Accordingly, we can minimize $\fnorm{\mF_m - \vw_m\vz_m^T}$ by taking $\vw_m\vz_m^T = s_m^{(1)} [\mU_m]_{:,0} [\mV_m]_{:,0}^H$ \cite{eckart36}. The complexity of computing the SVD of each rank-$2$ matrix $\mF_m$ is in $O(K)$ \cite{brand06}, so the overall complexity of the receiver is still in $O(KM\log KM)$. Moreover, we will show by simulation that this approximated MMSE receiver has favorable MSE and SER performance. 
Note that Theorem \ref{thm:mmselow} does not imply the non-existence of a low-complexity MMSE receiver in $O(KM\log KM)$ when both conditions (a) and (b) therein are not satisfied; it just states that an MMSE receiver cannot be implemented in the form shown in Fig. \ref{fig:CMRx2}. Whether an exact MMSE receiver can be implemented with low complexity remains an open question.

In summary, a low-complexity MMSE receiver implementation exists in an AWGN channel (as has been known). A less known condition for the existence of low-complexity MMSE receiver implementation is to employ a \emph{unitary GFDM matrix}. If one chooses not to use a unitary GFDM matrix, the approximated MMSE receiver can be used for a low-complexity implementation with suboptimal performance.

\subsection{Remarks on Soft-Output Demodulation}
In a receiver that applies soft-output demodulation, it is essential to have the knowledge of error variances $\sigma_{k,m}^2 \triangleq [\mR_e]_{k+mK}$, $\forall\; 0\leq k<K$, $0\leq m<M$, where $\mR_e = \e\{\ve\ve^H\}$ and $\ve=\hat{\vd}-\vd$.
It is worthy to note that low-complexity algorithms at the order $O(KM\log K)$ can be found to obtain {these} values, using characteristic matrix techniques presented above.
For the ZF receiver, using (\ref{eq:estsym}) and (\ref{eq:circdiag}), we can derive $\mR_e = N_0 (\mW_D\mA^{-H})^H \mD_C^{-1}\mD_C^{-H} (\mW_D\mA^{-H})$.
One may verify with some efforts, using (\ref{eq:cmimpl2}), that $\sigma^2_{k,m}$ is constant in $m$ for any $k$, and that the vector ${\boldsymbol\sigma} \triangleq [\sigma^2_{0,m} \sigma^2_{1,m} \cdots  \sigma^2_{K-1,m}]^T$ has the form
\begin{equation}
    {\boldsymbol\sigma} = \frac{N_0}{D} \mW_K \sum^{M-1}_{l=0}\left[ \diag(\vt_l) \mW_K^H \vr_l
    \right]\label{eq:sigmakmv}
\end{equation}
where
$\vr_l = [|C_l|^{-2} |C_{M+l}|^{-2} \cdots |C_{(K-1)M+l}|^{-2}]^T$
and $\vt_l^T = \sum^{K-1}_{p=0} [\bar{\mH}]_{p,l}^* [[\bar\mH]_{\modd{p}_K, l} [\bar\mH]_{\modd{p+1}_K, l}
\cdots [\bar\mH]_{\modd{p+K-1}_K, l}]$, with $\bar{\mH} = \bar{\mG}^{\circ -1}$.
Note that $\vt_l$ can be pre-computed, so the complexity for calculating (\ref{eq:sigmakmv}) is at the order $O(KM\log K)$.
For the MMSE receiver, assuming $\mA$ is unitary, we can similarly derive $\mR_e = E_S (\mI_D - (\mW_D\mA^{-H})^H \mD_C^H (\mD_C\mD_C^H + \gamma^{-1}\mI_D)^{-1} \mD_C (\mW_D\mA^{-H}))$.
Thus, we can derive that ${\boldsymbol\sigma}$ for the MMSE receiver can be expressed as in (\ref{eq:sigmakmv}) by changing the $k$th entry of $\vr_l$ from $|C_{kM+l}|^{-2}$ to $|C_{kM+l}|^2 / (|C_{kM+l}|^2 + \gamma^{-1})$, so the complexity for calculating the error variances is again at the order $O(KM\log K)$.

\begin{table*}[!t]
\scriptsize
\renewcommand{\arraystretch}{1.3}
\caption{Computational complexity of GFDM transceiver implementations under multipath channels}
\label{tbl:complexity}
\centering
\begin{tabular}{c|c|c|c}
\hline
Implementation & Transmitter & ZF receiver & MMSE receiver \\
\hline
\hline
OFDM & $\frac{1}{2}KM\log KM$ & $\frac{1}{2}KM\log KM+KM$ & $KM(\frac{1}{2}\log KM+1)$ \\
\hline
Direct & $K^2M^2$ & $K^2M^2+KM(\log KM+1)$ & $\frac{7}{3}K^3M^3 + 2K^2M^2$ \\
\hline
Frequency-domain \cite{michailow12b}& $KM(\frac{1}{2}\log KM^2+L_T)^a$ & $KM(\frac{1}{2}\log KM^2+L_R)+KM^b$ & Applicable only to AWGN channels \\
\hline
Frequency-convolution \cite{matthe16b} & $KM(\frac{1}{2}\log K+M)$ & $KM(\frac{1}{2}\log K+M)+KM(\log KM+1)$ & Applicable only to AWGN channels \\
\hline
Block-circularity \cite{farhang16} & $KM(\frac{1}{2}\log K+M)$ & $KM(\frac{1}{2}\log K+M)+KM(\log KM+1)$ & Applicable only to AWGN channels \\
\hline
Block-circularity \cite{farhang16}, power-of-2 $M$ & $KM(\frac{1}{2}\log KM^2+1)$ & $KM(\frac{1}{2}\log KM^2+1)+KM(\log KM+1)$ & Applicable only to AWGN channels \\
\hline
Zak-domain \cite{matthe15d} & Not applicable & Not applicable & $KM(\log M+6K+12M+4)^a$ \\
\hline
LU-decomposition \cite{matthe16a} & Not applicable & Not applicable & $KM(\frac{1}{2}\log KM+20M^2+22M)^a$ \\
\hline
Proposed Form 1 & $KM(\frac{1}{2} \log KM^2+1)$ & $KM(\frac{1}{2} \log KM^2+1)+KM(\log KM+1)$ & $KM(\frac{1}{2}\log K^3M^4+4)^c$ \\
\hline
Proposed Form 2 & $KM(\frac{1}{2}\log K^3M^2+1)$ & $KM(\frac{1}{2}\log K^3M^2+1)+KM$ & $KM(\frac{1}{2}\log K^3M^2+4)^c$ \\
\hline
\multicolumn{4}{l}{$^a$Assumption: The frequency-domain prototype transmit filter $\vg_f$ has only $L_T M$ or $2M$ nonzero entries, depending on the context.}\\
\multicolumn{4}{l}{$^b$Assumption: The frequency-domain prototype receive filter $\vh_f$ has only $L_R M$ nonzero entries.}\\
\multicolumn{4}{l}{$^c$Assumption: The prototype transmit filter is a CMCM filter.}
\end{tabular}
\end{table*}

\ignore
{

There are also three types of implementation for the receiver: direct implementation, frequency-domain implementation and characteristic-matrix-domain implementation. The drawback of the direct implementation is its high complexity $O(K^3M^3)$. Though the frequency-domain implementation has a lower complexity than the former, the MMSE receiver cannot be realized by this implementation in general. We again propose two forms of the characteristic-matrix-domain implementation. Both forms have the lowest complexity in $O(KM\log KM)$, with Form 2 has a bit lower complexity. Moreover, the MMSE receiver can be realized by this type of implementation if the transmitter matrix is unitary, or approximated by this type if the transmitter matrix is not unitary (up to a scale factor).

}

\section{Complexity Analysis} \label{sec:complexity}
The computational complexity of the proposed transceiver implementations in Sections \ref{sec:sysmod} and \ref{sec:receiver} is compared to that of several GFDM and conventional OFDM transceiver implementations.
As the case of AWGN channels has been well studied, we focus our complexity analysis on the case under multipath channels, which are more general and more practical.
For GFDM transmitters and ZF receivers, we include the frequency-domain implementation \cite{michailow12b} mentioned in Sections \ref{sec:sysmod} and \ref{sec:receiver}, the implementation proposed in \cite{matthe16b}, which is based on performing frequency-domain convolution in time domain as element-wise vector multiplication, and the implementation in \cite{farhang16
}, which is based on exploiting the block circularity of matrices involved in modulation and demodulation.
For GFDM MMSE receivers, we include the implementation in \cite{matthe15d}, which is based on calculating filter coefficients and filtering in the Zak domain, and the implementation in \cite{matthe16a}, which is based on simplifying the inversion of a band-diagonal matrix with LU decomposition. Since \cite{matthe16a} is for a multiple-antenna system, we calculate its complexity by reducing it to a single-antenna system. It is assumed in \cite{matthe15d, matthe16a} that the frequency-domain prototype transmit filter $\vg_f$ has only $2M$ nonzero entries, so the complexity formulae for the MMSE receivers in \cite{matthe15d, matthe16a} cannot be used for all general prototype filters.
We also compare to direct implementations, where the matrix multiplications and inverses in (\ref{eq:xad}), (\ref{eq:estsym}), and (\ref{eq:estsymmmse}) are implemented directly.
The comparison is based on the number of complex multiplications (CMs) required to transmit or receive $KM$ symbols, as shown in Table \ref{tbl:complexity}.
For a fair comparison, the same block size $KM$ as GFDM is used for OFDM \cite{lin10}.
To obtain the complexity formulae, we assume that a $p$-point DFT \cite{duhamel1990} and the inversion of a $p\times p$ matrix based on Gaussian elimination \cite{strang09} take $\frac{p}{2}\log p$ and $p^3/3$ CMs, respectively, for any positive integer $p$, where the base of the logarithm is $2$.
The prototype filters for all implementations are assumed to take complex values.
Since the prototype filter in \cite{farhang16} is assumed to be real-valued, we extend their results to the case of complex-valued filters.

As depicted in Fig. \ref{fig:CMTx1}, the proposed Form-1 transmitter implementation involves four steps: $M$ sets of $K$-point inverse-DFTs (IDFTs), $K$ sets of $M$-point DFTs, element-wise multiplication with a $K\times M$ matrix, and $K$ sets of $M$-point IDFTs. These result in $M\frac{K}{2}\log K + K\frac{M}{2}\log M + KM + K\frac{M}{2}\log M = KM(\frac{1}{2} \log KM^2+1)$ CMs.
Similarly, we can derive the complexity formulae for the proposed Form-2 transmitter (\ref{eq:cmimpl2}), Form-1 receiver (\ref{eq:CMRx1}), and Form-2 receiver (\ref{eq:CMRx2}) as described in Table \ref{tbl:complexity}.
If CMCM filters are used, MMSE receivers can also be implemented based on (\ref{eq:CMRx1}) and (\ref{eq:CMRx2}) by replacing each diagonal entry $C_l$ in the matrix $\mD_C$ with $C_l+(\gamma \xi_G C_l^*)^{-1}$, where $\xi_G$ is the energy of the transmitter matrix, and $\gamma=E_S/N_0$ is the SNR.
In view of the number of CMs described in Table \ref{tbl:complexity}, we recommend using the Form-1 implementation for transmitters and Form-2 implementation for receivers.
For the frequency-domain implementation \cite{michailow12b}, the parameter $L_T\leq K$ is the number of subcarriers spanned by the frequency-domain prototype transmit filter (i.e., $\vg_f$ has only $L_TM$ nonzero entries), and $L_R\leq K$ is the number of subcarriers spanned by the frequency-domain prototype receive filter.
It was stated in \cite{farhang16} that the complexity of their implementation can be reduced when $M$ is a power of two. The reduced complexity is listed separately in Table \ref{tbl:complexity}.
For a fair comparison, frequency-domain one-tap equalization $\mW_D^H \mD_C^{-1} \mW_D$, taking $KM(\log KM+1)$ CMs, as in (\ref{eq:CMRx1}) or in (7) of \cite{farhang16} is used for all GFDM ZF receivers except for the proposed Form-2 receiver and the implementation in \cite{michailow12b}, in which, due to cancellation of a pair of DFT and IDFT, only $KM$ additional CMs are needed.

%
%
\begin{figure}[!t]
\centering
\includegraphics[width=3.1in]{\figpath/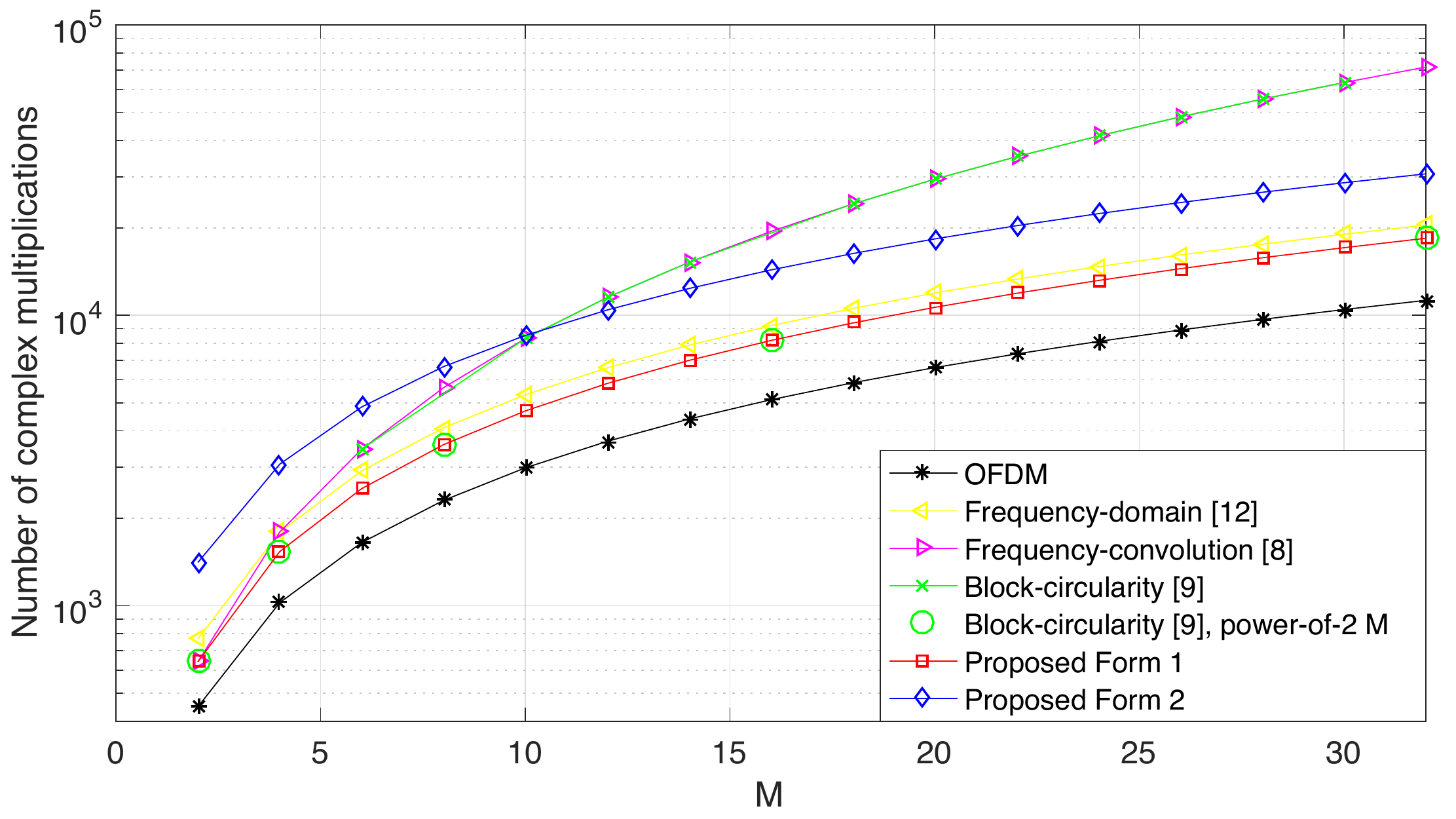}
\caption{Complexity of GFDM transmitter implementations.}
\label{fig:complexitytx}
\end{figure}

\begin{figure}[!t]
\centering
\includegraphics[width=3.1in]{\figpath/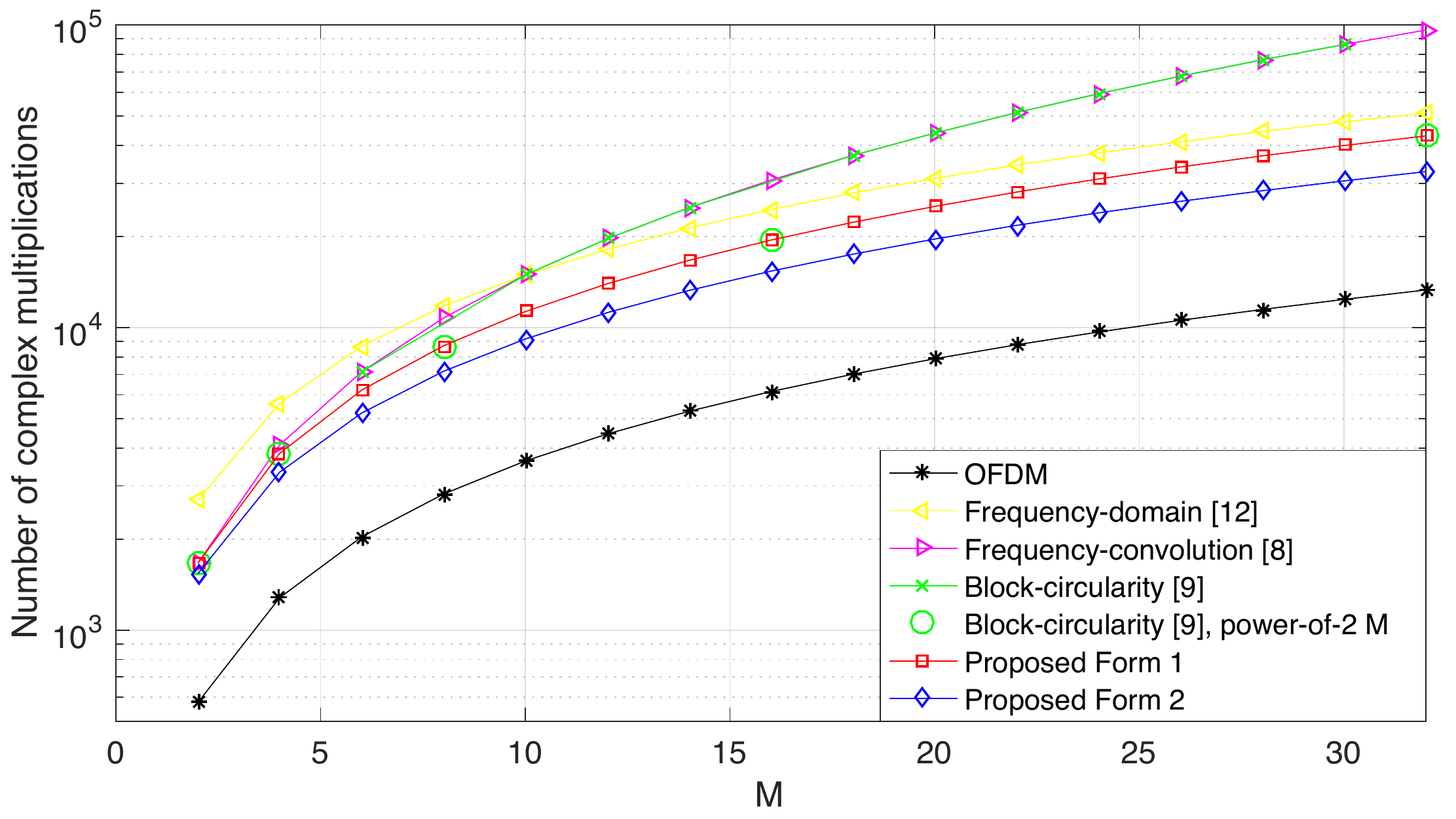}
\caption{Complexity of GFDM ZF receiver implementations.}
\label{fig:complexityzfrx}
\end{figure}

\begin{figure}[!t]
\centering
\includegraphics[width=3.1in]{\figpath/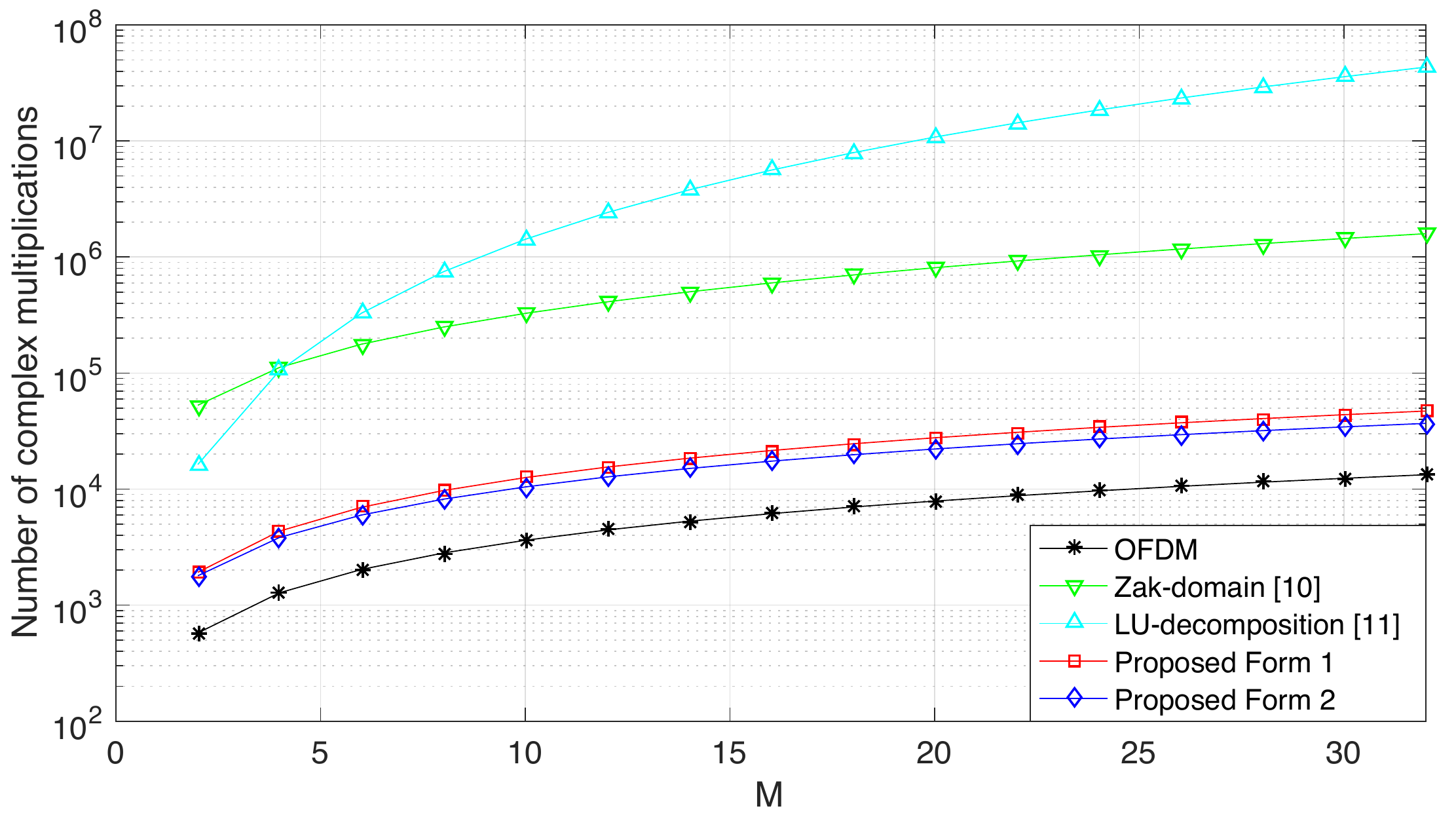}
\caption{Complexity of GFDM MMSE receiver implementations.}
\label{fig:complexitymmserx}
\end{figure}

The complexity formulae in Table \ref{tbl:complexity} are evaluated and plotted for $K=64$ subcarriers with respect to different values of number of subsymbols $M$.
The complexity of the transmitter implementations is shown in Fig. \ref{fig:complexitytx}.
As suggested in \cite{michailow12b}, $L_T=2$ is chosen for calculating the complexity of the frequency-domain implementation.
According to Fig. \ref{fig:complexitytx}, the number of CMs required by the proposed Form-1 transmitter is the least among all GFDM transmitters, and is only about $1.5$ times as much as that required by the OFDM transmitter.
The complexity of the frequency-domain implementation \cite{michailow12b}, under the assumption that $L_T$ is as small as $2$, is around $1.1$ to $1.2$ times the complexity of the proposed Form-1 transmitter.
The complexity of the implementation in \cite{matthe16b} and the one in \cite{farhang16} is even over $3$ times the complexity of the proposed Form-1 transmitter when $M$ is relatively large.
(The complexity of the implementation in \cite{michailow12b} would be higher than that of the one in \cite{matthe16b} if $L_T=K$ for general filters.)
The reduced complexity of the implementation in \cite{farhang16} when $M$ is a power of two, nevertheless, coincides that of the proposed Form-1 transmitter.

The complexity of the ZF receiver implementations is shown in Fig. \ref{fig:complexityzfrx}.
Based on the suggestion in \cite{michailow14}, $L_R=16$ is chosen for the frequency-domain ZF receiver implementation \cite{michailow12b}.
According to Fig. \ref{fig:complexityzfrx}, the number of CMs required by the proposed Form-2 ZF receiver is the least among all GFDM ZF receivers, and is only about $2.5$ times as much as that required by the OFDM ZF receiver.
The complexity of the frequency-domain implementation \cite{michailow12b} is around $1.6$ to $1.8$ times the complexity of the proposed Form-2 ZF receiver.
The complexity of the implementation in \cite{matthe16b} and the one in \cite{farhang16} is even nearly $3$ times the complexity of the proposed Form-2 ZF receiver when $M$ is relatively large.
The reduced complexity of the implementation in \cite{farhang16} is still around $1.1$ to $1.3$ times the complexity of the proposed Form-2 ZF receiver when $M$ is a power of two.

The complexity of the MMSE receiver implementations is shown in Fig. \ref{fig:complexitymmserx}.
We see in Fig. \ref{fig:complexitymmserx} that the number of CMs required by the proposed Form-2 MMSE receiver is the least among all GFDM MMSE receivers, and is only about $2.8$ times as much as that required by the OFDM MMSE receiver.
Compared to the implementations in \cite{matthe15d, matthe16a}, complexity reduction of around 2 to 3 orders of magnitude can be achieved by the proposed Form-2 MMSE receiver because the complexity of the proposed implementation is linearithmic while that of the one in \cite{matthe15d} is quadratic with the numbers of both subsymbols $M$ and subcarriers $K$, and that of the one in \cite{matthe16a} is even cubic with the number of subsymbols.

In summary, with the use of the proposed implementations, significant complexity reduction can be obtained for receivers, while moderate complexity reduction is also obtained for transmitters.
Note that direct implementations are not shown in Figs. \ref{fig:complexitytx}, \ref{fig:complexityzfrx}, and \ref{fig:complexitymmserx} since they demand extremely large numbers of CMs. For example, when $K=64$ and $M=16$, they require about two orders of magnitude more CMs than the proposed implementations do for a transmitter or ZF receiver, and about five orders of magnitude more CMs than the proposed implementations do for an MMSE receiver.

\section{Prototype Filter Design} \label{sec:mmse}

We propose in this section to design optimal prototype filters in terms of minimizing the receiver MSE before considering the OOB radiation performance. Due to the one-to-one relation between the prototype transmit filter $\vg$ and the characteristic matrix $\mG$ in Lemma \ref{lmm:gfdmpoly}(a), the design of the characteristic matrix is essentially equivalent to the prototype filter design. We address the problem mainly from the perspective of the characteristic matrix, which yields many insights.

The receiver MSE is formally defined as follows. Denote the error variance on the $k$th subcarrier and $m$th subsymbol after demodulation as
\begin{equation}
    \sigma_{k,m}^2=\e\left\{|[\hat{\vd}-\vd]_{k+mK}|^2\right\} \label{eq:errorvar}
\end{equation}
for $k=0,1,\dots,K-1$ and $m=0,1,\dots,M-1$, where $\hat{\vd}$ is defined as in (\ref{eq:estsym}) or (\ref{eq:estsymmmse}) if the ZF or MMSE receiver is used, respectively. The expectation is taken on both the noise and channel distributions. Define the receiver MSE $\sigma^2$ as
\begin{equation}
    \sigma^2=\frac{1}{D}\sum_{k=0}^{K-1}\sum_{m=0}^{M-1}\sigma_{k,m}^2. \label{eq:msedef}
\end{equation}
Our goal is to identify the optimal $K\times M$ characteristic matrix $\mG$ of a $D\times D$ GFDM matrix $\mA$ that minimizes the receiver MSE $\sigma^2$ as defined in (\ref{eq:msedef}) under the following scenarios:
\begin{enumerate}
    \item the ZF receiver over the AWGN channel;\label{pbm:msezfawgn}
    \item the ZF receiver over (statistical) multipath channels;\label{pbm:msezfst}
    \item the MMSE receiver over the AWGN channel;\label{pbm:msemmseawgn}
    \item the MMSE receiver over (statistical) multipath channels,\label{pbm:msemmsest}
\end{enumerate}
which we call Problems 1-4. We fix $\xi_G$, which equals $\|\vg\|^2$ by Lemma \ref{lmm:gfdmpoly}(d), as a normalization of the energy of the prototype filter.

\subsection{Optimization Results for ZF Receivers}




The solutions to Problems \ref{pbm:msezfawgn} and \ref{pbm:msezfst} are identified in the following theorem, with some additional requirements introduced for Problem \ref{pbm:msezfst}.

\vspace{0.5em}

\begin{theorem} \label{thm:mmseprob12}
(a) Under the ZF receiver over the AWGN channel, a prototype transmit filter $\vg$ minimizes MSE $\sigma^2$ if and only if it is a CMCM filter. The corresponding minimum MSE is $\sigma^2_{min}=N_0/\xi_G$.\\
(b) Under the ZF receiver over any statistical channel such that the channel frequency response $C_l$ satisfies $\e\{1/|C_{l}|^2\}$ being a finite constant $\beta,\forall\;0\leq l<D$, a prototype transmit filter $\vg$ minimizes MSE $\sigma^2$ if and only if it is a CMCM filter. The corresponding minimum MSE is $\sigma^2_{min}=\beta N_0/\xi_G$.
\end{theorem}
\begin{IEEEproof}
(a) By (\ref{eq:estsym}) with $\mC=\mI_D$ and Theorem \ref{thm:prop}(c),
\begin{IEEEeqnarray}{rCl}
    \sigma_{k,m}^2=\e\left\{\left|[\mA^{-1}\vq]_{k+mK}\right|^2\right\}=\xi_HN_0, \label{eq:unifzfawgn}
\end{IEEEeqnarray}
$\forall\; 0\leq k<K, 0\leq m<M$, where $\xi_H$ is the energy of $\mA^{-H}$.
Then, the statement follows from (\ref{eq:msedef}) and the inequality $\xi_G\xi_H\geq 1$, which is shown below. By Theorem \ref{thm:prop}(b), $\xi_H = \sum_{k=0}^{K-1}\sum_{l=0}^{M-1}(1/(D|[\mG]_{k,l}|^2))$.
Then, $\xi_G\xi_H\geq 1$ follows from the Cauchy-Schwarz inequality, 
\begin{IEEEeqnarray}{rCl}
    \left[\sum_{k=0}^{K-1}\sum_{l=0}^{M-1}|[\mG]_{k,l}|^2\right]\left[\sum_{k=0}^{K-1}\sum_{l=0}^{M-1}\frac{1}{|[\mG]_{k,l}|^2}\right] \geq (KM)^2, \IEEEeqnarraynumspace
\end{IEEEeqnarray}
where the equality holds if and only if $|[\mG]_{k,l}|$ is a constant in both $k$ and $l$. 
The expression for $\sigma^2_{min}$ follows from (\ref{eq:unifzfawgn}) and the condition for the equality to hold for $\xi_G\xi_H\geq 1$.
\\
(b) Taking the expectation of (\ref{eq:sigmakmv}) and noting that $\e\{{\vr_l}\} = \beta \mathbf{1}_K$, we can derive that
\begin{equation}
    \sigma_{k,m}^2=\beta\|\bar{\mH}\|^2 N_0/D, \forall\; 0\leq k< K, 0\leq l< M. \label{eq:unifzfstt}
\end{equation}
Thus, we have $\sigma^2 = \beta \|\bar{\mH}\|^2 N_0/D = \beta\xi_H N_0$, and the result follows from $\xi_G\xi_H\geq 1$ as proved in (a).
\end{IEEEproof}

\vspace{0.5em}

Note that in Theorem \ref{thm:mmseprob12}(b) (i.e., solution to Problem \ref{pbm:msezfst}), $\e\{1/|C_{l}|^2\}$ is required to be a finite constant $\forall\;0\leq l<D$. Requiring them to be finite is a necessary condition for the receiver MSE $\sigma^2$ to also be finite, and is an inherent limitation of a ZF receiver since $\sigma^2 \propto \e\{1/|C_{l}|^2\}, \forall\;0\leq l<D$ in this case. Besides, we require them to be a constant so that there remains some sort of symmetry as we move from the AWGN channel to statistical channels.

After considering the statistical case for Problem \ref{pbm:msezfst}, we now evaluate the static case. Specifically, we consider a deterministic multipath channel, or a slow fading channel such that obtaining perfect channel state information at the transmitter (CSI-T) is practical. The solution is as follows.

\vspace{0.5em}

\begin{theorem} \label{thm:mmsefs}
Under the ZF receiver over any (static) multipath channel $C_l$ such that $C_l\neq 0, \forall\; 0\leq l<D$, a prototype transmit filter $\vg$ minimizes MSE $\sigma^2$ if and only if $|[\mG]_{k,l}|^2/\sqrt{\alpha_l}$ is a constant in both $k$ and $l$, where $\alpha_l=\sum_{r=0}^{K-1}1/(|C_{l+rM}|^2)$. The corresponding minimum MSE is $\sigma^2_{min}=(\sum_{l=0}^{M-1}\sqrt{\alpha_{l}})^2N_0/(KM^2\xi_G)$.
\end{theorem}
\begin{IEEEproof}
See Appendix \ref{adx:mmsefs}.
\end{IEEEproof}

\vspace{0.5em}

The proposed filters in Theorem \ref{thm:mmsefs} are optimal in terms of minimizing MSE, but they require CSI-T and are less applicable than the CMCM filters derived under statistical channels in Theorem \ref{thm:mmseprob12}(b).

\subsection{Optimization Results for MMSE Receivers}
The solution to Problem \ref{pbm:msemmseawgn} is given by the following theorem, whose proof is similar to that of Theorem \ref{thm:mmseprob12}(a).

\vspace{0.5em}

\begin{theorem} \label{thm:mmseprob3}
Under the MMSE receiver over the AWGN channel, a prototype transmit filter $\vg$ minimizes MSE $\sigma^2$ if and only if it is a CMCM filter. The corresponding minimum MSE is $\sigma^2_{min}=E_S/(\gamma\xi_G+1)$.
\end{theorem}
\begin{IEEEproof}
See Appendix \ref{adx:mmseineq}.
\end{IEEEproof}

\vspace{0.5em}

Observing that each of the solutions to Problems \ref{pbm:msezfawgn}, \ref{pbm:msezfst}, and \ref{pbm:msemmseawgn} is a CMCM filter, we make the following conjecture that the solution to Problem \ref{pbm:msemmsest}, under the assumption of identically distributed $C_l$, $\forall\;0\leq l<D$, is also a CMCM filter.

\vspace{0.5em}

\emph{Hypothesis 1:} Under the MMSE receiver over any statistical channel such that the channel frequency response $C_l$ are identically distributed $\forall\;0\leq l<D$, a prototype transmit filter $\vg$ minimizes MSE $\sigma^2$ if and only if it is a CMCM filter. The corresponding minimum MSE is $\sigma^2_{min}=\e\{E_S/(\gamma\xi_G|C_0|^2+1)\}$.

\vspace{0.5em}

In Hypothesis 1, the assumption of identically distributed $C_l$, $\forall\;0\leq l<D$ is practical since many realistic channels, such as Rayleigh fading channels \cite{haykin01}, have identically distributed $C_l$.
Note that we do not require each $\e\{1/|C_{l}|^2\}$ to be finite because an MMSE receiver does not suffer from this limitation. While a mathematical proof for Hypothesis 1 is unavailable now because the inverse of $\mC\mA\mA^H\mC^H + \gamma^{-1}\mI_D$ in (\ref{eq:mmmserxmat}) cannot be readily simplified (one may consider properties of block circulant matrices for the simplification in the future), numerical results in Section \ref{sec:sim} verify that this hypothesis tends to be correct.

The solutions to all the four problems provide criteria for the prototype transmit filter $\vg$ to minimize the MSE under various types of channels and receivers. Since some degrees of freedom (i.e., $\angle [\mG]_{k,l}$) remain in all the solutions, minimizing the OOB radiation with respect to $\vg$ under the derived criteria would be a suitable research direction for future studies.

\subsection{Comparison of Prototype Filter Candidates}
Considering the optimization results, we find it natural to categorize GFDM prototype filters into two classes: The first class comprises CMCM filters, corresponding to the class of unitary GFDM matrices (up to a scale factor), and the second class comprises non-CMCM filters, corresponding to the class of non-unitary GFDM matrices. The first class is advantageous in minimizing the receiver MSE, whereas the second class suffers from the noise enhancement effect \cite{matthe14b, michailow12a, han16arxiv}.

The RC, RRC, Xia \cite{xia97}, and Gaussian pulses \cite{matthe14a}, adopted by many previous studies, fall into the class of non-CMCM filters. GFDM systems using these filters are non-orthogonal \cite{matthe14a}. 
In fact, since RC and RRC filters are even-symmetric, i.e., $[\vg]_n=[\vg]_{D-n}$ for $n=1,2,\dots,D-1$, their GFDM matrices are singular when $K,M$ are both even integers.
This can be proved by using (\ref{eq:chrmtxdef}) to show that the corresponding characteristic matrix $\mG$ satisfies $[\mG]_{\frac{K}{2}, \frac{M}{2}}=0$ and using Theorem \ref{thm:prop}(a) (see also \cite{matthe14b}, which also observed this point using Gabor analysis).
Thus, to avoid MSE and SER performance degradation, we would not set $D=KM$ as a power of 2 for GFDM systems using RC and RRC filters. By contrast, the simulation results in this paper show that if the prototype transmit filter is not even-symmetric, both $K$ and $M$ being even does not prevent a GFDM system from exhibiting good MSE and SER performance. There is also no constraint on $K$ and $M$ in Theorem \ref{thm:unitary} for GFDM matrices to be unitary.

\ignore
{
One subtlety is about the construction of the RC and RRC filters in our context. The RC and RRC filters are originally of infinite length in the continuous-time domain, so we perform truncation and sampling when adopting them as GFDM prototype transmit filters. Due to the truncation, the DFTs of the RC and RRC prototypes are not exactly the samples of the continuous frequency response of the original RC and RRC filters. For the same reason, in our context the Dirichlet pulse is approximately but not exactly a special case of the RC and RRC filters with a zero roll-off factor.
}

The class of CMCM prototype filters were less common in previous studies. Yet, their existence implies that noise enhancement is not always a problem of GFDM. As a simple example of CMCM filters, consider the GFDM matrix whose phase-shifted characteristic matrix $\bar{\mG}$ satisfies
\begin{equation}
    [\bar{\mG}]_{k,l}=1, \forall\; 0\leq k<K, 0\leq l<M. \label{eq:freqshifteddirichlet}
\end{equation}
The corresponding frequency-domain prototype filter is $[\vg_f]_l=\sqrt{K}\sum_{k=0}^{M-1}\delta_{lk}$, $l=0, 1, \dots, D-1$.
In fact, it is a frequency-shifted version of the Dirichlet pulse \cite{matthe14a}. The Dirichlet pulse is defined by a perfect rect function in the frequency domain with the width of $M$ frequency bins located around the DC bin\footnote{Although this definition is only clear for an odd $M$, we give a reasonable extension for an even $M$.}. In other words, by defining $\cX_1=\{0, 1, \dots, \lfloor\frac{M-1}{2}\rfloor\}$ and $\cX_2=\{D-\lceil\frac{M-1}{2}\rceil, D-\lceil\frac{M-1}{2}\rceil+1, \dots, D-1\}$, we can express the frequency-domain prototype filter as
\begin{equation}
    [\vg_f]_l=\sqrt{K}\sum_{k\in\cX_1\cup\cX_2}\delta_{lk},l=0, 1, \dots, D-1. \label{eq:dirichlet}
\end{equation}
It is also a CMCM filter, and its corresponding GFDM matrix is unitary, as shown in the following corollary.

\vspace{0.5em}

\begin{corollary} \label{crl:dirichlet}
The GFDM matrix for the Dirichlet pulse is unitary.
\end{corollary}
\begin{IEEEproof}
By Lemma \ref{lmm:gfdmpoly}(b) and (\ref{eq:dirichlet}), we can derive that the phase-shifted characteristic matrix $\bar{\mG}$ satisfies $\forall\; 0\leq k<K$,
\begin{IEEEeqnarray}{rCl}
    [\bar{\mG}]_{k,l}&=&\left\{\begin{array}{lr}1, & 0\leq l < \lceil M/2\rceil \\
    e^{-j2\pi k/K}, & \lceil M/2\rceil\leq l <M
    \end{array}\right.,\IEEEeqnarraynumspace
\end{IEEEeqnarray}
for the Dirichlet pulse. Thus, it is a CMCM filter, and by Theorem \ref{thm:unitary}, the corresponding GFDM matrix is unitary.
\end{IEEEproof}

\vspace{0.5em}

The Dirichlet pulse (\ref{eq:dirichlet}) instead of its frequency-shifted version (corresponding to (\ref{eq:freqshifteddirichlet})) will be used in the simulations because its passband is centered at the DC bin.

As another example of CMCM filters, we propose the \emph{modified Dirichlet pulse}, defined by the frequency response
\begin{IEEEeqnarray}{rCl}
    [\vg_f]_l&=&\sqrt{K}e^{j\pi\frac{l}{D}} \sum_{k\in\cX_1}\delta_{lk} + \sqrt{K}e^{j\pi\frac{l-D}{D}}\sum_{k\in\cX_2}\delta_{lk},\IEEEeqnarraynumspace \label{eq:designgl}
\end{IEEEeqnarray}
$l=0, 1, \dots, D-1$.
The phase-shifted characteristic matrix $\bar{\mG}$ for the filter satisfies
\begin{IEEEeqnarray}{rCl}
    [\bar{\mG}]_{k,l}&=&\left\{\begin{array}{lr}e^{j\pi l/D}, & 0\leq l < \lceil M/2\rceil \\
    e^{j\pi(-2kM+(l-M))/D}, & \lceil M/2\rceil\leq l <M
    \end{array}\right.,\IEEEeqnarraynumspace 
\end{IEEEeqnarray}
$\forall\; 0\leq k< K$, so it is a CMCM filter. Later we will explain why this filter may have lower OOB radiation than does the Dirichlet pulse and verify this through simulation.

%
%
\begin{figure}[!t]
\centering
\includegraphics[width=3.6in]{\figpath/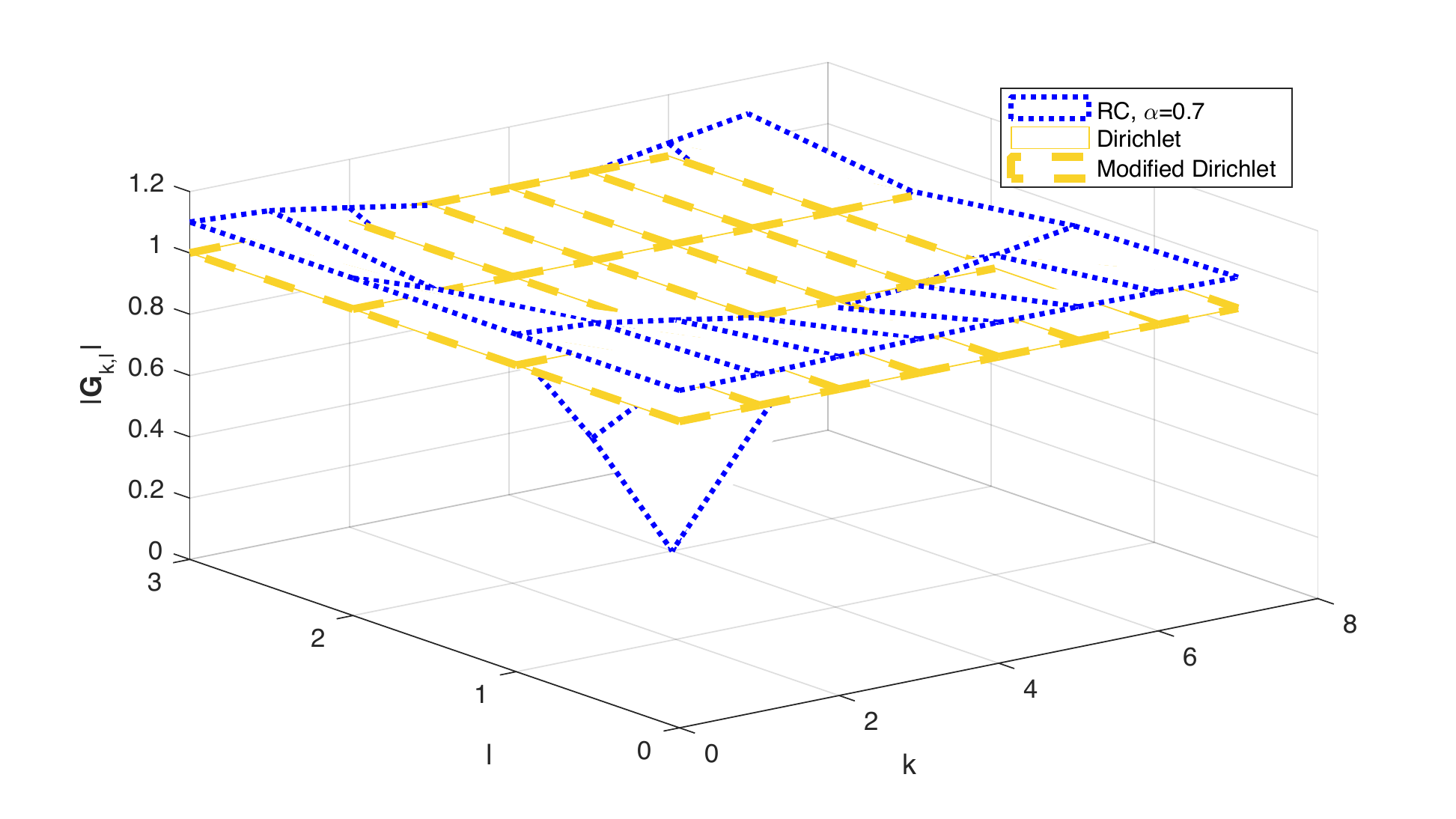}
\caption{Magnitudes of entries in characteristics matrices $\mG$ for several GFDM prototype filters when $K=8,M=4$.}
\label{fig:pf}
\end{figure}

The absolute values of the entries of characteristic matrices for the RC filter with roll-off factor $\alpha=0.7$, the Dirichlet pulse, and the modified Dirichlet pulse are compared in Fig. \ref{fig:pf}. Note the zero for the RC filter (which makes the corresponding GFDM matrix singular). However, the other two are advantageous because they have constant magnitudes in the characteristic matrix, making both of them unitary.

\section{Power Spectral Density and OOB Leakage}\label{sec:oob}
In this section, which serves as an aid for simulation later, we define the OOB leakage $O$ as a performance measure for the OOB radiation of transmit signals. To evaluate $O$ for GFDM, we first address the power spectral density (PSD) of GFDM signals. We derive an analytical PSD expression encompassing an interpolation filter used in a D/A converter. This approach conforms to the practical realization of modern digital-signal-processing-based communication systems \cite{lin10}.

The GFDM digital baseband transmit signal $x[n]$ is described as in (\ref{eq:dbts}). The analog baseband transmit signal $x_a(t)$ is obtained by passing $x[n]$ through a D/A converter with a sampling interval $T_s$ and an interpolation filter $p(t)$, i.e., $x_a(t)=\sum_{n=-\infty}^{\infty}x[n]p(t-nT_s)$.
The PSD of $x_a(t)$ is defined as $S_a(f)=\lim_{T\to\infty}\e\{\frac{1}{2T}|\int_{-T}^T x_a(t) e^{-j2\pi ft}\:\mathrm{d}t|^2\}$ \cite{haykin01}.
Let $P(f)=\int_{-\infty}^{\infty} p(t) e^{-j2\pi ft}\:\mathrm{d}t$ be the Fourier transform of $p(t)$, and $G_m(e^{j\omega})=\sum_{n=-\infty}^{\infty} g_m[n] e^{-j\omega n}$ be the discrete-time Fourier transform of $g_m[n]$, where $g_m[n]$ is defined in (\ref{eq:gmn}).
Assuming the data symbols are zero-mean and i.i.d. with symbol energy $E_S$, we can derive that
\begin{IEEEeqnarray}{rCl}
    S_a(f)=\frac{E_S|P(f)|^2}{D'T_s}\sum_{k\in\cK}\sum_{m\in\cM}\left|G_m\left(e^{j2\pi\left(fT_s-\frac{k}{K}\right)}\right)\right|^2.\IEEEeqnarraynumspace\label{eq:psd}
\end{IEEEeqnarray}
A special case that leads to a simple expression of $G_m(e^{j\omega})$ is $L=0$. When $L=0$, we derive that
\begin{equation}
    G_m(e^{j\omega})=\sum_{l=0}^{D-1}[\vg_f]_l e^{-j2\pi \frac{lm}{M}}\sinc_D\left(\omega_ l\right)e^{-j\omega_l\frac{D-1}{2}},\label{eq:gmf}
\end{equation}
where $\vg_f$ is the frequency-domain prototype transmit filter, $\omega_l=\omega-(2\pi l/D)$, and
\begin{equation}
    \sinc_D(x)=\left\{\begin{array}{ll}(-1)^{k(D-1)}, & x=2\pi k, k\in\bZ \\
    \frac{\sin(Dx/2)}{D\sin(x/2)}, & \text{otherwise}
    \end{array}\right.
\end{equation}
is the periodic sinc function. Substituting (\ref{eq:gmf}) into (\ref{eq:psd}), we can express the PSD with $\vg_f$, which enables designing the PSD in terms of the frequency-domain prototype transmit filter.

Here we explain why the modified Dirichlet pulse exhibits lower OOB radiation than does the Dirichlet pulse. Taking the absolute value of (\ref{eq:gmf}) and setting $m=0$ yields $|G_0(e^{j\omega})| = |\sum_{l=0}^{D-1}[\vg_f]_l\sinc_D(\omega_l)e^{j\pi l\frac{D-1}{D}}|$. Since $\sinc_D(x)$ alternates between positive and negative values as $x$ crosses nonzero integer multiples of $2\pi/D$, $\sinc_D(x)+e^{j\phi}\sinc_D(x-2\pi/D)$ with $\phi=\pi$ can be viewed as the extreme case of "constructive interference" for the tails of the periodic sinc functions. Thus, as $e^{j\pi l\frac{D-1}{D}}$ and the factor $e^{j\pi l/D}$ introduced in (\ref{eq:designgl}) combine to form $e^{j\pi l}$, the modified Dirichlet pulse exhibits lower OOB radiation than does the Dirichlet pulse under the scenario that the $0$th subsymbol is used as a guard symbol. In other words, we allocate as much OOB energy as possible on the discarded subsymbol.

To characterize the OOB radiation, we define the OOB leakage \cite{matthe14a} as
\begin{equation}
    O=\frac{|\cB_I|}{|\cB_O|}\cdot\frac{\int_{f\in\cB_O}S_a(f)\:\mathrm{d}f}{\int_{f\in\cB_I}S_a(f)\:\mathrm{d}f}, \label{eq:oobdef}
\end{equation}
In (\ref{eq:oobdef}), $\cB_I$ and $\cB_O$ are the set of frequencies considered in-band and out of band, respectively, and $|\cB_I|$ and $|\cB_O|$ denote the lengths of the corresponding intervals. Recall that $\cK$ is the set of subcarrier indices actually used.
The nominal frequencies of the subcarriers in $\cK$ lie in $\cB_I$, several guard subcarriers are used between $\cB_I$ and $\cB_O$, and $\cB_O$ is reserved for the use of other users.

\begin{figure*}[!t]
\centering
\subfloat[MSE for ZF-DFERF, $K=8,M=5$.]{\includegraphics[width=2.3in, height=1.9in]{\figpath/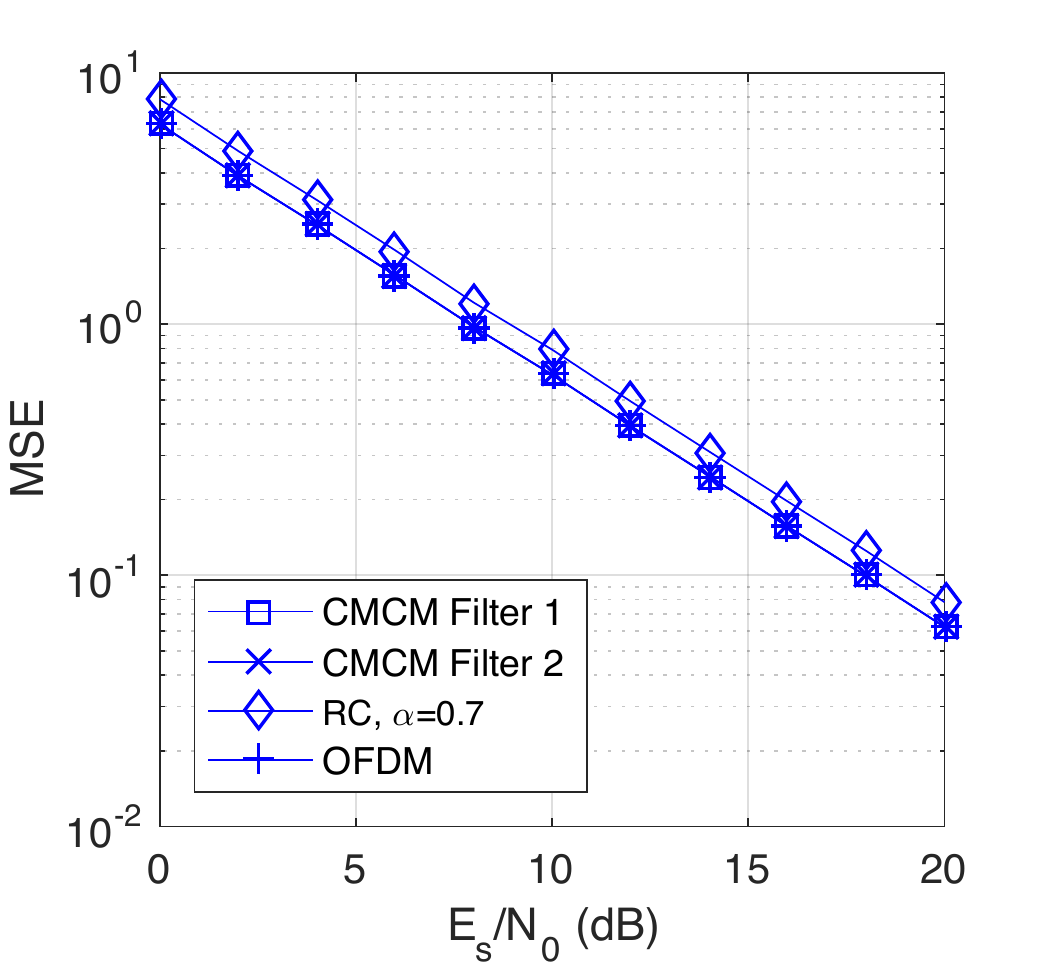}
\label{MSE_ZFRx_DFERayleighCH_K8M5_SNR0to20}}
\subfloat[MSE for ZF-DFERF, $K=8,M=4$.]{\includegraphics[width=2.3in, height=1.9in]{\figpath/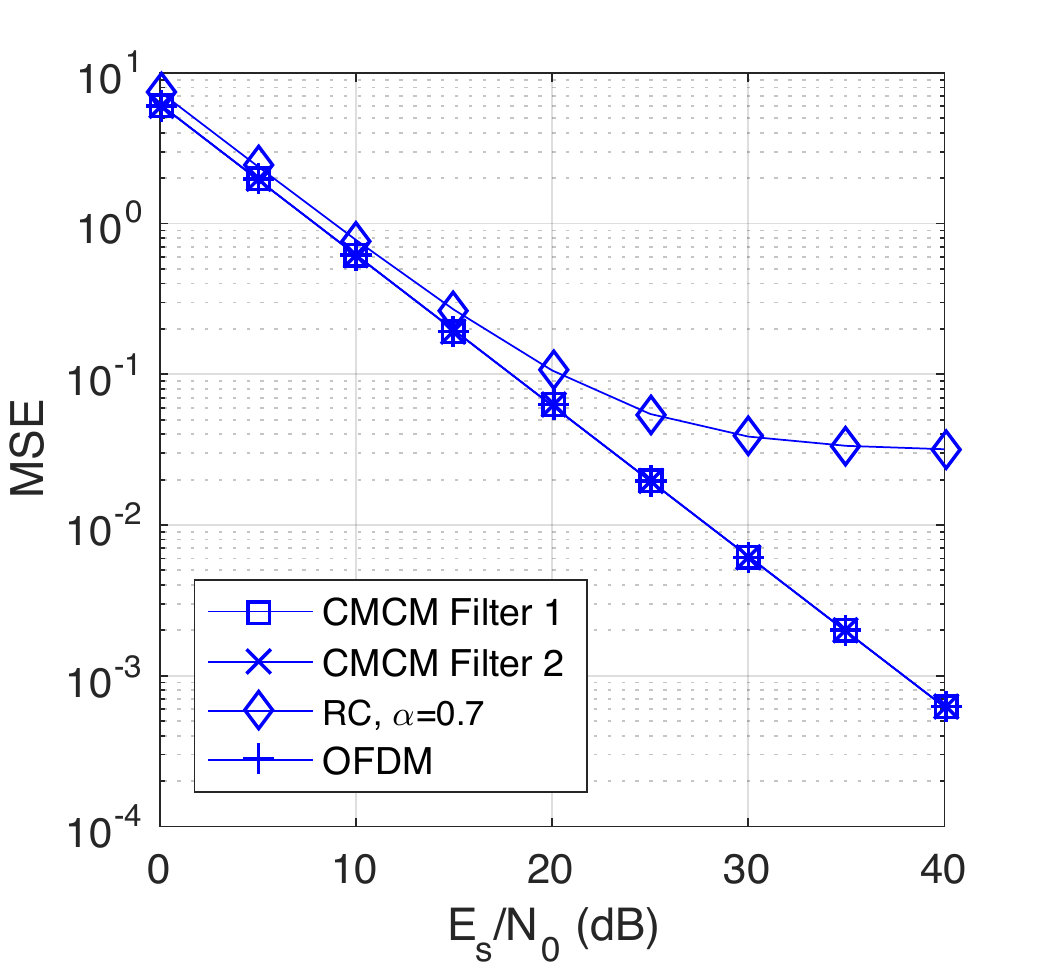}
\label{MSE_ZFRx_DFERayleighCH_K8M4_SNR0to40}}
\subfloat[MSE for MMSE-RF, $K=32,M=16$.]{\includegraphics[width=2.3in, height=1.9in]{\figpath/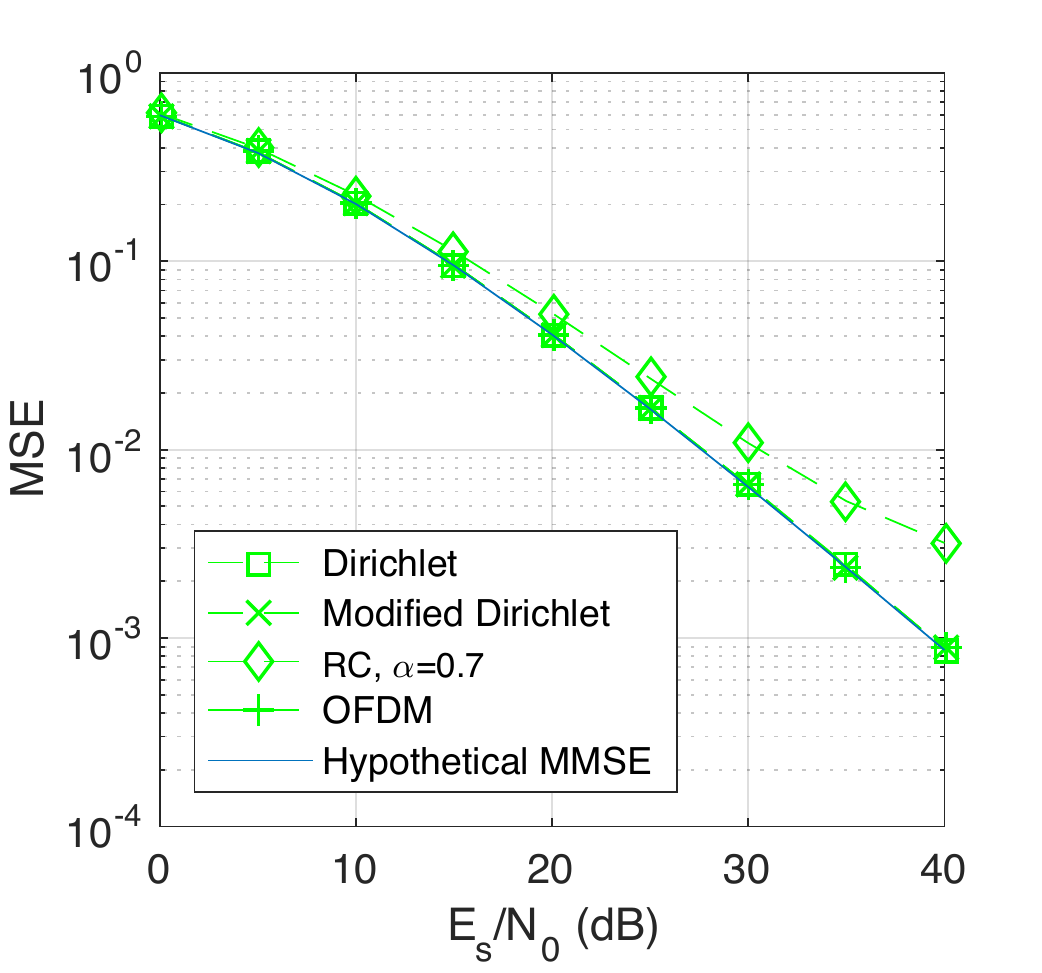}
\label{MSE_MMSERx_RayleighCH_K8M4_SNR0to30}}
\caption{MSE for GFDM ZF receiver over a deep-fade-excluded Rayleigh fading (DFERF) channel, MMSE receiver over the Rayleigh fading (RF) channel, and the corresponding OFDM receivers.} \label{fig:statistical}
\end{figure*}

Finally, note that in (\ref{eq:dbts}), the sets $\cM$ and $\cK$ are not required to be $\cM=\{0, 1, \dots, M-1\}$ or $\cK=\{0, 1, \dots, K-1\}$. This means some guard symbols or guard subcarriers can be used. GFDM is proposed to exhibit low OOB radiation. This advantage is particularly significant if some guard symbols and guard subcarriers are used \cite{matthe14a}. To address the effects of nulling some data symbols on the MSE or SER performance, we present the following corollary.

\vspace{0.5em}

\begin{corollary}\label{crl:uniform}
For any GFDM systems, $\sigma_{k,m}^2$ is a constant in both $k$ and $m$ for each of the scenarios: the ZF receiver over the AWGN channel, the MMSE receiver over the AWGN channel, and the ZF receiver over any statistical channel such that the channel frequency response $C_l$ satisfies $\e\{1/|C_{l}|^2\}$ being a finite constant $\beta,\forall\;0\leq l<D$.
\end{corollary}
\begin{IEEEproof}
See Appendix \ref{adx:uniform}.
\end{IEEEproof}

\vspace{0.5em}

The main idea of the proof is that the equal-norm property in Theorem \ref{thm:prop}(c) implies equal noise enhancement for each subcarrier and subsymbol. These results imply that we can just null the data symbols leading to the highest OOB radiation without considering the MSE performance of each subcarrier or subsymbol, as demonstrated in the simulation.


\vspace{-0.1cm}
\section{Simulation Results} \label{sec:sim}
\vspace{-0.1cm}

In this section, we provide numerical examples to compare the derived optimal prototype transmit filters, especially the CMCM filters, with the conventional RC and RRC filters, in terms of receiver MSE, SER, and OOB leakage.

\subsection{MSE and SER Performance}\label{ssc:mseser}
The MSE and SER performance is evaluated through Monte-Carlo simulation with 10000 blocks for each prototype filter under each of the following six cases:
\begin{enumerate}
    \item ZF-DFERF: the ZF receiver over a deep-fade-excluded Rayleigh fading channel;\label{sim:zfdferay}
    \item MMSE-RF: the MMSE receiver over the Rayleigh fading channel;\label{sim:mmseray}
    \item AMMSE-RF: the low-rank approximated MMSE receiver (see Sec. \ref{ssc:ammse}) over the Rayleigh fading channel;\label{sim:ammseray}
    \item ZF-AWGN: the ZF receiver over the AWGN channel;\label{sim:zfawgn}
    \item MMSE-AWGN: the MMSE receiver over the AWGN channel;\label{sim:mmseawgn}
    \item ZF-MP: the ZF receiver over a (static) multipath channel.\label{sim:zffs}
\end{enumerate}
We use $(K, M)=(8, 4)$, $(8, 5)$, or $(32, 16)$ for GFDM, and $K=32$, $40$, or $512$, $M=1$ for OFDM (OFDM is a special case of GFDM using a rectangular window as the prototype transmit filter) so that GFDM and OFDM have the same block size $D=KM$.
For the used Rayleigh fading channels, $c[n]$ is independent circularly symmetric complex Gaussian with variance $N^{(c)}_n$, and two kinds of power delay profiles \cite{lin10} are used. For cases when $(K, M)=(8, 4)$ or $(8, 5)$, we use $N_n^{(c)}=(0.64)^n$ for $0\leq n<D/4$ and $N_n^{(c)}=0$ for $D/4\leq n<D$. For cases when $(K, M)=(32, 16)$, we use $N_n^{(c)}=0, -1, -2, -3, -8, -17.2, -20.8$ dB for $n=0, 3, 7, 9, 11, 19, 41$, respectively, and $N_n^{(c)}=0$ otherwise, which is derived from the LTE Extended Pedestrian A model \cite{sesia11}.
In Case ZF-DFERF, the channel is derived from the Rayleigh fading channel by excluding all channel realizations leading to a tap gain $|C_l|$ smaller than -30 dB for some subcarrier $l$ from the channel pool. This exclusion results in finite $\e\{1/|C_l|^2\}$, and is practical since transmission is given up when deep fades occur in real communication.
The (static) multipath channel (in case ZF-MP) is composed of four taps: $-0.1518+j0.6475, 0.2701+j0.3063, 0.5703+j0.0767, -0.0900+j0.2274$.
Finally, for MMSE receivers (Cases \ref{sim:mmseray}, \ref{sim:ammseray}, and \ref{sim:mmseawgn}), unbiased estimates are used for symbol detection \cite{lin10}.

\begin{figure*}[!t]
\centering
\subfloat[MSE.]{\includegraphics[width=2.3in]{\figpath/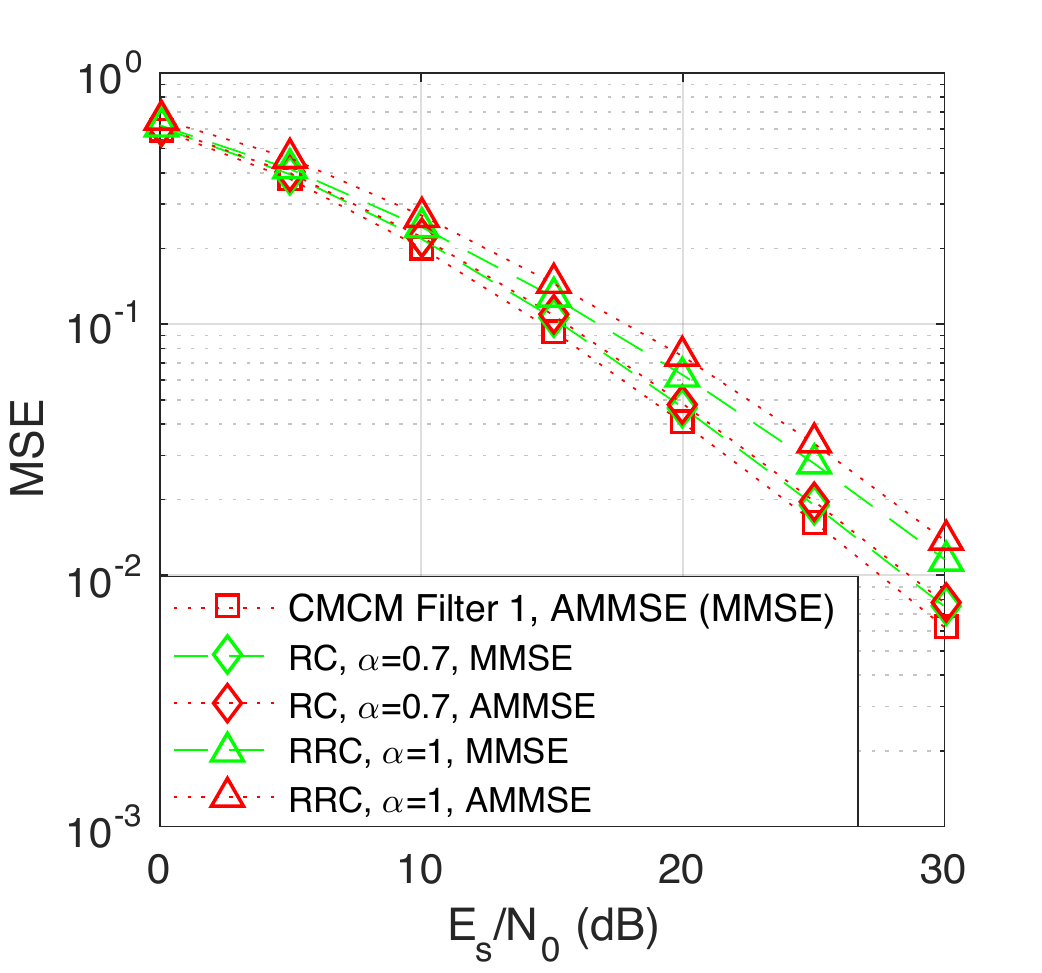}
\label{MSE_MMSERxApr_RayleighCH_K8M5_SNR0to30}}
\subfloat[SER.]{\includegraphics[width=2.3in]{\figpath/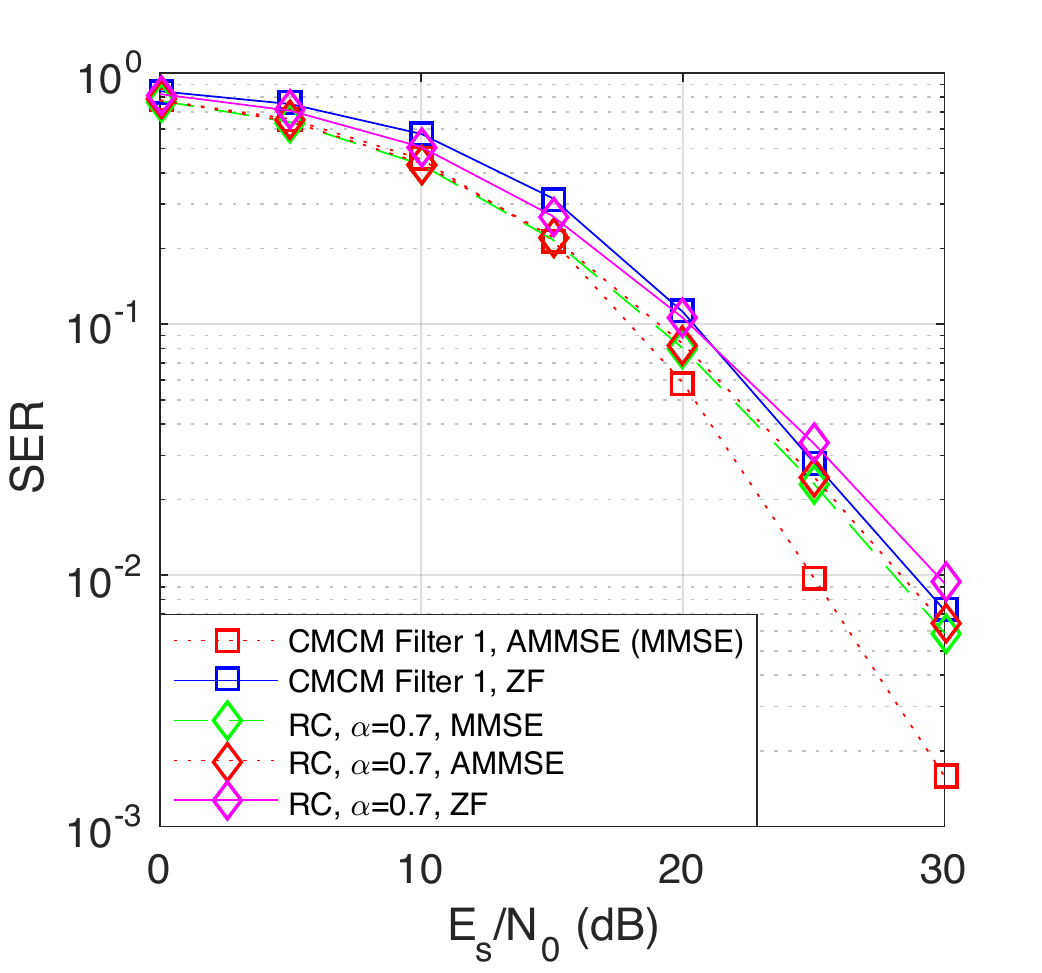}
\label{SER_MMSERxApr_RayleighCH_K8M5_SNR0to30_noRRC}}
\subfloat[SER.]{\includegraphics[width=2.3in]{\figpath/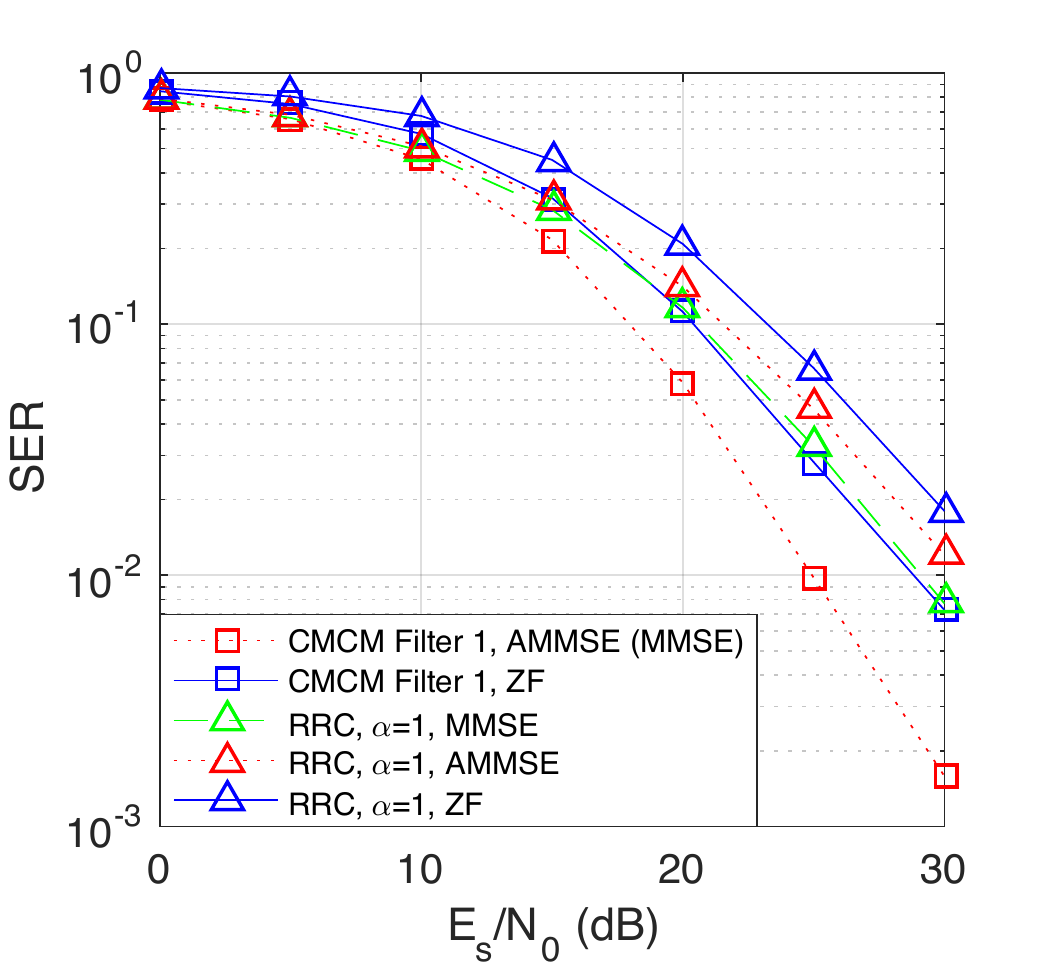}
\label{SER_MMSERxApr_RayleighCH_K8M5_SNR0to30_noRC}}
\caption{MSE and SER for GFDM approximated MMSE (AMMSE) receiver over the Rayleigh fading channel (compared to the corresponding ZF and MMSE receivers). $K=8, M=5$.} \label{fig:AMMSE}
\end{figure*}

The modulation is 16QAM, the symbol energy is $E_S=1$, and the energy of the GFDM transmitter matrices is $\xi_G=1$. The CP length is $L=D/4$.
The prototype transmit filters used for GFDM include an RC filter with roll-off factor $\alpha=0.7$ and an RRC filter with roll-off factor $\alpha=1$.
To demonstrate that MSE performance is not affected by phases of entries of constant-magnitude characteristic matrices, we use CMCM filters with arbitrarily chosen phases in most of the simulation cases (except for ZF-MP).
Specifically, we use CMCM Filters 1 and 2 with characteristic matrices $\mG_1$ and $\mG_2$, respectively, 
with the phases $\angle\mG_1$ and $\angle\mG_2$ being arbitrarily selected and listed as follows. For systems with $K=8,M=4$, $\angle\mG_1$ and $\angle\mG_2$ are set as
\begin{equation}{\fontsize{7}{6}\selectfont
    \angle\mG_1=\begin{bmatrix}
        0.75 & 2.50 & -1.09 & -1.98 \\
        -2.95 & 0.16 & 1.29 & 1.59 \\
        -2.10 & 0.59 & 3.12 & -0.31 \\
        0.53 & 3.04 & 0.28 & -1.11 \\
        1.58 & 1.37 & -3.02 & -1.80 \\
        -3.11 & 1.05 & 0.47 & -0.73 \\
        0.78 & -1.88 & 0.85 & -2.24 \\
        1.57 & -2.83 & -0.56 & 2.81
    \end{bmatrix},\label{eq:ph1k8m4}
}\end{equation}
\begin{equation}{\fontsize{7}{6}\selectfont
    \angle\mG_2=\begin{bmatrix}
        -0.31 & -3.11 & 0.82 & -1.04 \\
        -1.70 & 2.53 & -0.29 & 0.71 \\
        -2.49 & 2.19 & -2.69 & -1.55 \\
        -1.44 & -0.77 & -2.06 & 0.19 \\
        0.23 & -1.00 & 0.31 & 0.48 \\
        0.95 & -1.50 & 2.26 & 0.09 \\
        0.21 & -1.03 & 0.76 & 0.57 \\
        2.17 & 1.79 & -2.15 & 1.88
    \end{bmatrix}.\label{eq:ph2k8m4}
}\end{equation}
For systems with $K=8,M=5$, $\angle\mG_1$ and $\angle\mG_2$ are set as
\begin{equation}{\fontsize{7}{6}\selectfont
    \angle\mG_1=\begin{bmatrix}
        0.62 & -0.40 & -1.36 & -2.16 & -1.94 \\
        -1.30 & -2.65 & 2.78 & -2.95 & 2.17 \\
        1.01 & 0.07 & 2.86 & 2.92 & -0.60 \\
        1.75 & 2.09 & 1.59 & 0.48 & -1.89 \\
        1.55 & -1.83 & -0.11 & -3.01 & -0.57 \\
        0.27 & -1.21 & -2.81 & 0.37 & -2.27 \\
        -1.48 & 0.46 & 2.58 & 2.72 & 0.44 \\
        1.23 & -0.31 & 1.19 & 0.06 & -0.35
    \end{bmatrix},\label{eq:ph1k8m5}
}\end{equation}
\begin{equation}{\fontsize{7}{6}\selectfont
    \angle\mG_2=\begin{bmatrix}
        -2.89 & -1.87 & -2.40 & -3.02 & -1.22 \\
        0.73 & 2.22 & -2.79 & 3.08 & 3.04 \\
        0.90 & -2.14 & -1.51 & -2.13 & -1.69 \\
        -2.42 & -2.99 & -1.16 & -0.08 & -0.63 \\
        -1.94 & -2.57 & 2.22 & 1.17 & 2.89 \\
        1.33 & 1.10 & -2.51 & -1.44 & 1.36 \\
        -3.06 & -3.05 & -2.54 & -3.09 & 0.36 \\
        0.53 & 0.22 & 2.88 & -2.08 & 0.54
    \end{bmatrix}.
}\end{equation}
For systems with $K=32,M=16$, $\angle\mG_1$ and $\angle\mG_2$ are set in ways such that CMCM Filters 1 and 2 are the Dirichlet pulse (\ref{eq:dirichlet}) and modified Dirichlet pulse (\ref{eq:designgl}), respectively.
For Case ZF-MP, we also use two filters proposed in Theorem \ref{thm:mmsefs}, with the phases of the characteristic matrices $\angle\mG_1$ and $\angle\mG_2$ again arbitrarily set as in (\ref{eq:ph1k8m4}) and (\ref{eq:ph2k8m4}) for the case of $K=8,M=4$.
For cases where both $K$ and $M$ are even, the ZF receiver for an RC or RRC filter does not exist, so we use instead the pseudo-inverse of the GFDM matrix.

\begin{figure*}[!t]
\centering
\subfloat[MSE for ZF-AWGN.]{\includegraphics[width=2.3in]{\figpath/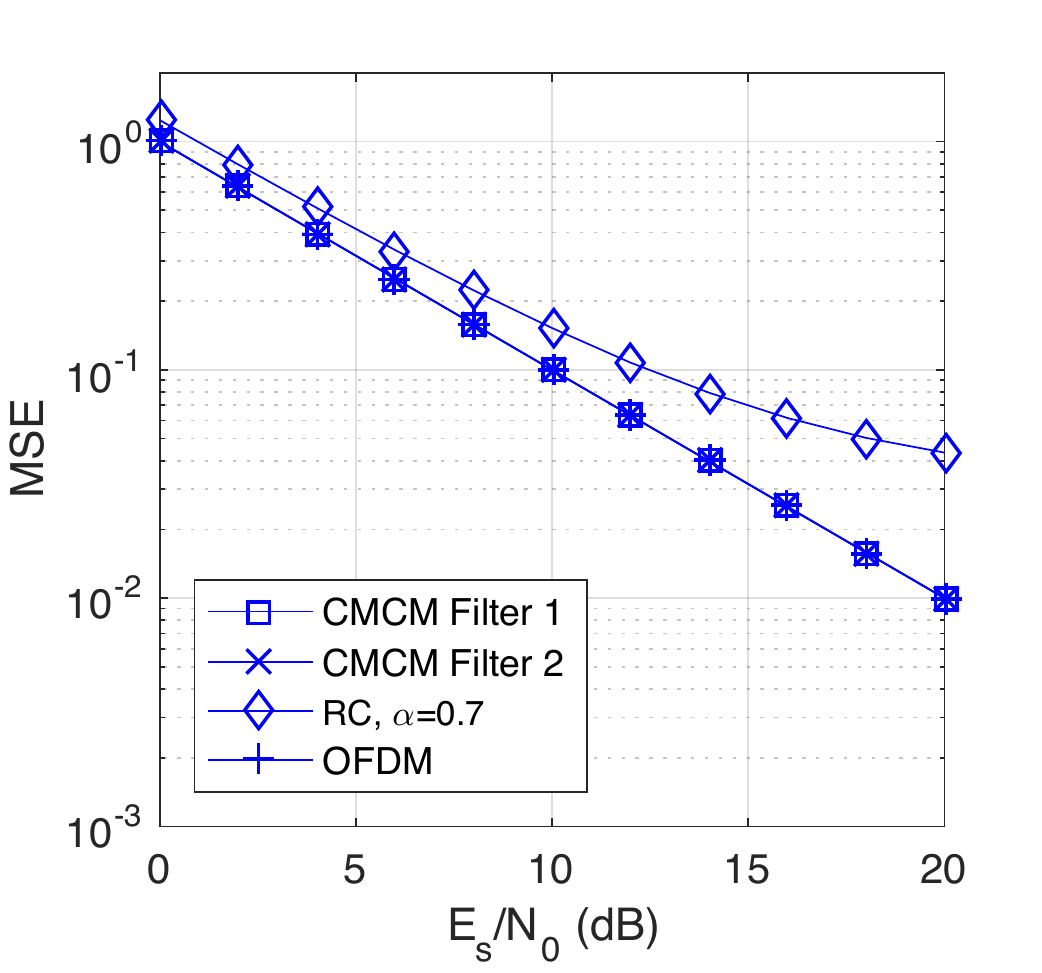}
\label{MSE_ZFRx_AWGNCH_K8M4_SNR0to20}}
\subfloat[MSE for MMSE-AWGN.]{\includegraphics[width=2.3in]{\figpath/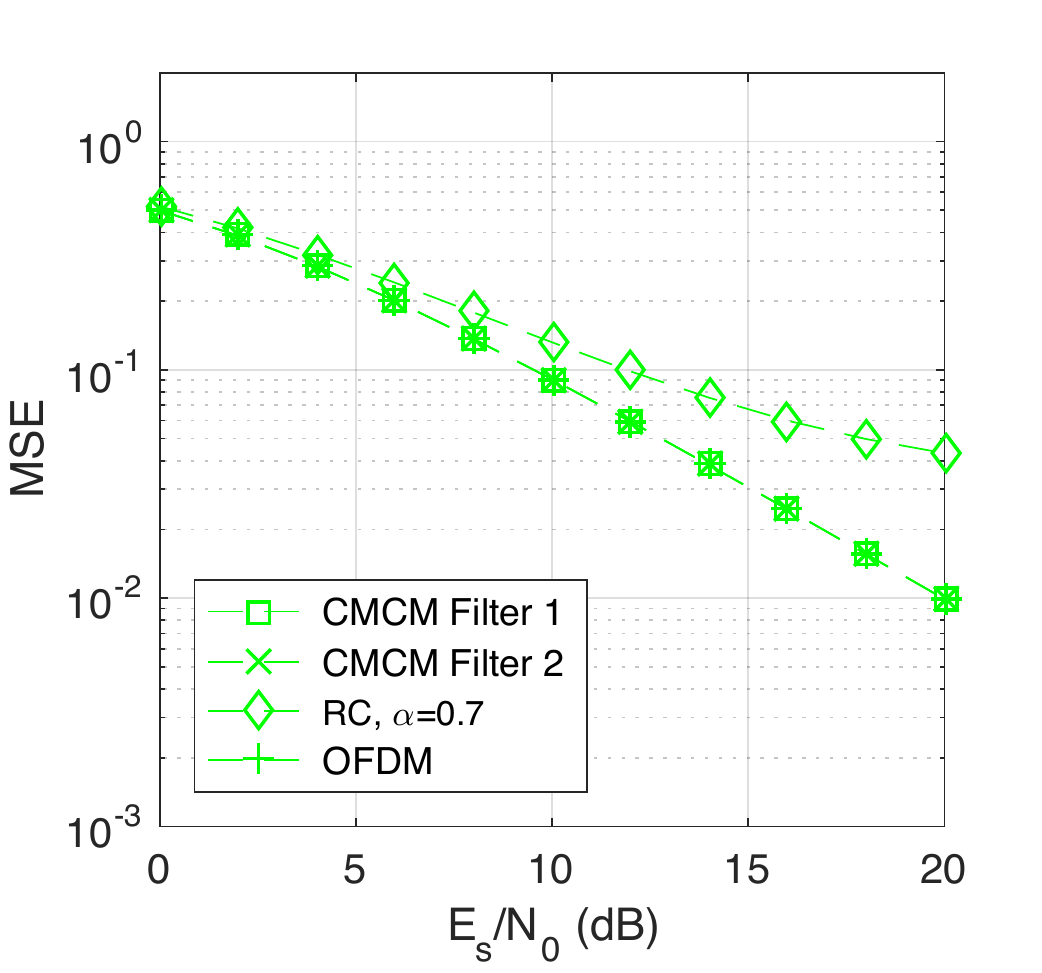}
\label{MSE_MMSERx_AWGNCH_K8M4_SNR0to20}}
\subfloat[MSE for ZF-MP.]{\includegraphics[width=2.3in]{\figpath/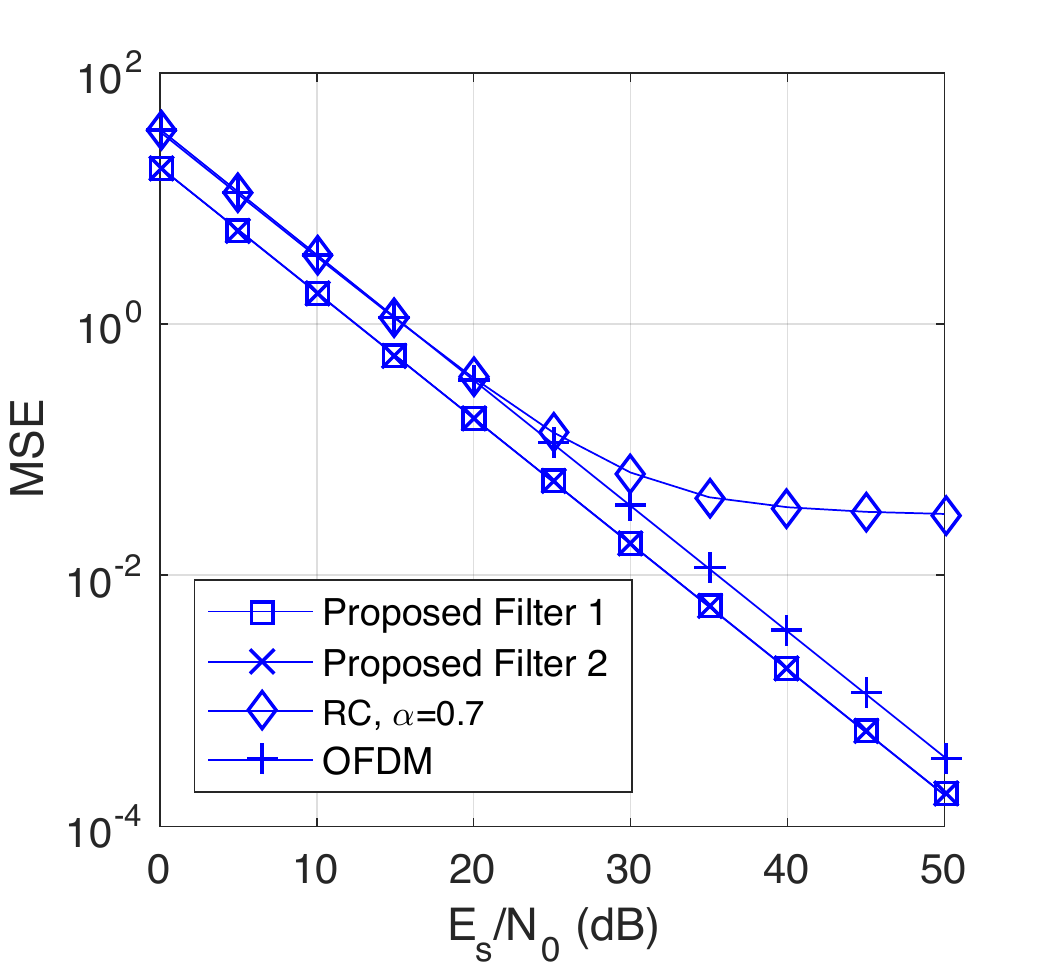}
\label{MSE_ZFRx_FSCH_K8M4_SNR0to40}}
\caption{MSE for GFDM ZF receiver over the AWGN channel, MMSE receiver over the AWGN channel, ZF receiver over a (static) multipath (MP) channel, and the corresponding OFDM receivers. $K=8, M=4$.} \label{fig:static}
\end{figure*}

\begin{figure}[!t]
\centering
\includegraphics[width=3.4in]{\figpath/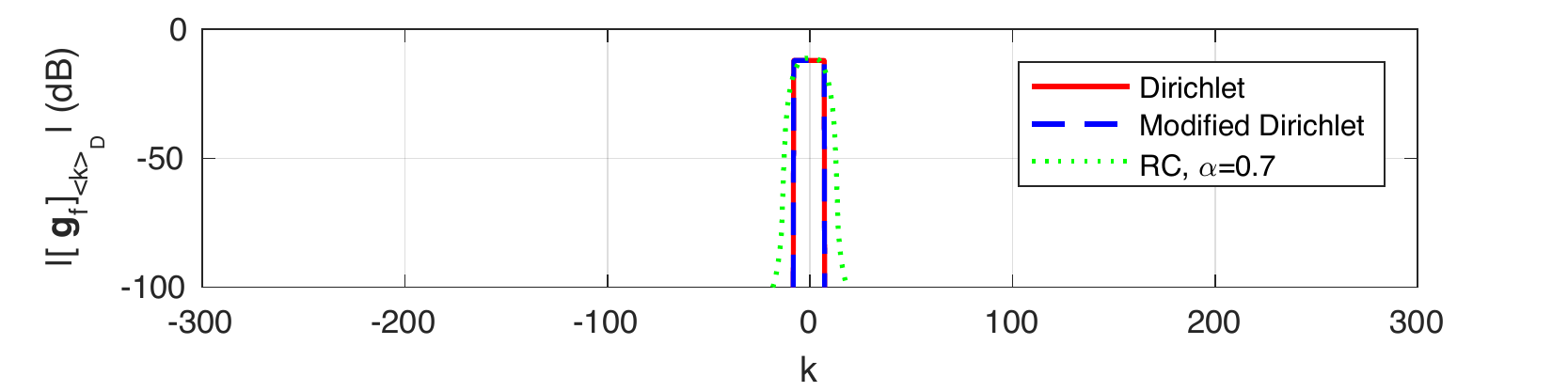}
\caption{Magnitude response of the frequency-domain prototype filter $\vg_f$.} \label{fig_gf}
\end{figure}

\begin{figure}[!t]
\centering
\includegraphics[width=3.4in]{\figpath/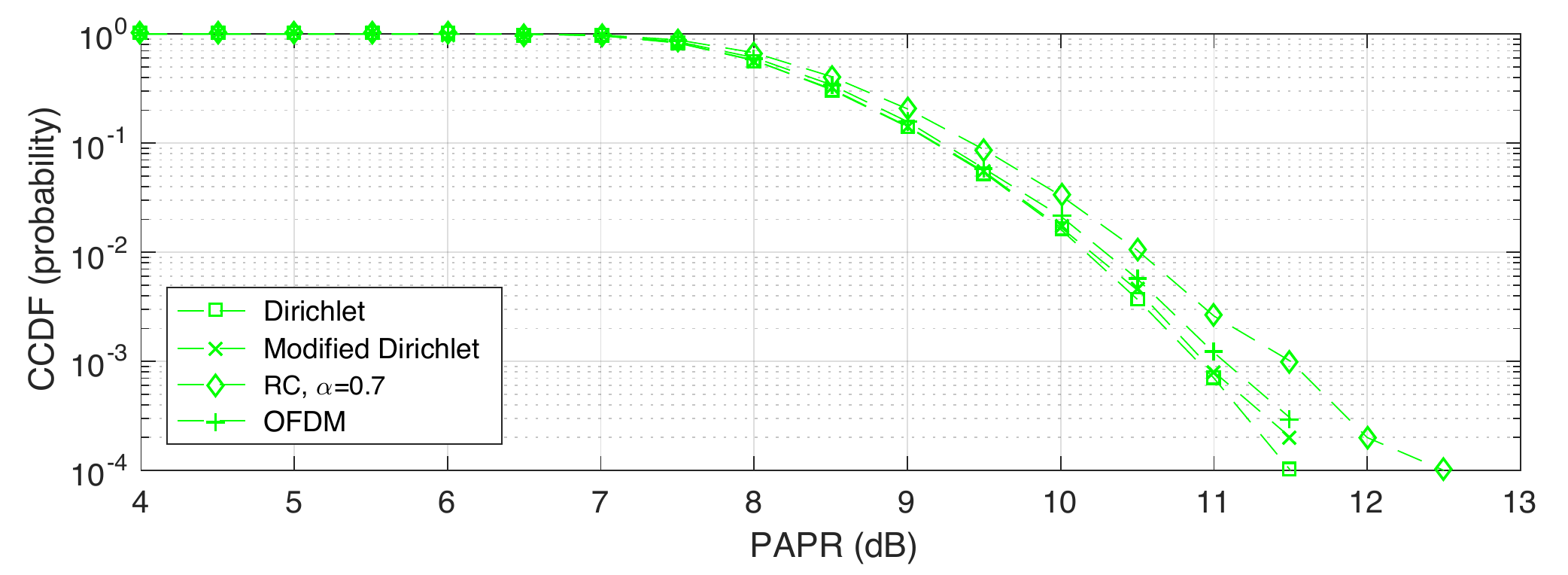}
\caption{CCDF of PAPR for MMSE-RF. $K=32, M=16$.} \label{fig_papr}
\end{figure}

Fig. \ref{fig:statistical} shows the simulation results under statistical channels.
We first consider the case $K=8, M=5$.
Fig. \subref*{MSE_ZFRx_DFERayleighCH_K8M5_SNR0to20} verifies the MMSE property of CMCM filters under the ZF receiver over the deep-fade-excluded Rayleigh fading channel, as stated in Theorem \ref{thm:mmseprob12}(b).
The CMCM filters are better than the RC filter, and essentially the same as OFDM in terms of MSE performance.
Turning to the case $K=8, M=4$, we see similar results in Fig. \subref*{MSE_ZFRx_DFERayleighCH_K8M4_SNR0to40}. Yet, the RC filter performs even worse due to the singularity of its transmitter matrix, whereas the CMCM filters do not have such degradation. Finally, similar results are again observed in Fig. \subref*{MSE_MMSERx_RayleighCH_K8M4_SNR0to30} for the case of the MMSE receiver over the Rayleigh fading channel. Meanwhile, the MSEs of the CMCM filters correspond to the hypothetical minimum MSE in Hypothesis 1. These imply that Hypothesis 1 tends to be correct.
Note that the channels used for the ZF and MMSE receivers have different statistics, so the results in Figs. \subref*{MSE_ZFRx_DFERayleighCH_K8M4_SNR0to40} and \subref*{MSE_MMSERx_RayleighCH_K8M4_SNR0to30} cannot be compared directly.

The performance of our proposed low-complexity approximated MMSE receiver is evaluated in Fig. \ref{fig:AMMSE}.
The SER performance is shown in two subfigures to make the curves clear.
As indicated by Theorem \ref{thm:mmselow}, the AMMSE and MMSE receivers for a CMCM filter are identical.
By contrast, Fig. \ref{fig:AMMSE} show that the MSE and SER performance is degraded due to approximation for the RC and RRC filters, particularly the RRC filter.
In this case, $\xi_H=1$, $1.08$, and $1.25$ for the CMCM, RC, and RRC filters, respectively.
Together with the results in Fig. \ref{fig:AMMSE}, this implies that higher nonuniformity of $|[\mG]_{k,l}|$, $\forall\; 0\leq k< K, 0\leq l< M$ engenders more errors in the approximation process.
However, in Figs. \subref*{SER_MMSERxApr_RayleighCH_K8M5_SNR0to30_noRRC} and \subref*{SER_MMSERxApr_RayleighCH_K8M5_SNR0to30_noRC}, all AMMSE receivers show significant performance improvements over their ZF receiver counterparts, and exhibit SERs that are the same as or only slightly higher than do the MMSE receiver counterparts.
(MSEs of ZF receivers are infinite and thus not shown in Fig. \subref*{MSE_MMSERxApr_RayleighCH_K8M5_SNR0to30}.)
Besides, the complexity of the MMSE receivers (directly implemented as in (\ref{eq:mmmserxmat2})) is $O(K^3M^3)$, whereas that of the AMMSE receivers is $O(KM\log KM)$.
These show that the AMMSE receiver is a good compromise between complexity and performance.

Fig. \ref{fig:static} shows the simulation results under static channels, including the AWGN channel.
Fig. \subref*{MSE_ZFRx_AWGNCH_K8M4_SNR0to20} shows that the CMCM filters are better than the RC filter, and essentially the same as OFDM in terms of MSE performance.
It verifies that the CMCM filters are the prototype filters that minimize receiver MSE under the ZF receiver over the AWGN channel, as stated in Theorem \ref{thm:mmseprob12}(a).
Similar results can be observed in Fig. \subref*{MSE_MMSERx_AWGNCH_K8M4_SNR0to20}.
It verifies that the CMCM filters are the prototype filters that minimize receiver MSE under the MMSE receiver over the AWGN channel, as stated in Theorem \ref{thm:mmseprob3}.
Fig. \subref*{MSE_ZFRx_FSCH_K8M4_SNR0to40} verifies the MMSE property of the proposed filters under the ZF receiver over (static) multipath channels as stated in Theorem \ref{thm:mmsefs}.
The advantages of the proposed filters in this case come from the use of CSI-T, whereas the RC filter and the prototype filter of OFDM, i.e., the rectangular window, are predefined and are not designed according to CSI-T.

\begin{figure}[!t]
\centering
\includegraphics[width=3.6in]{\figpath/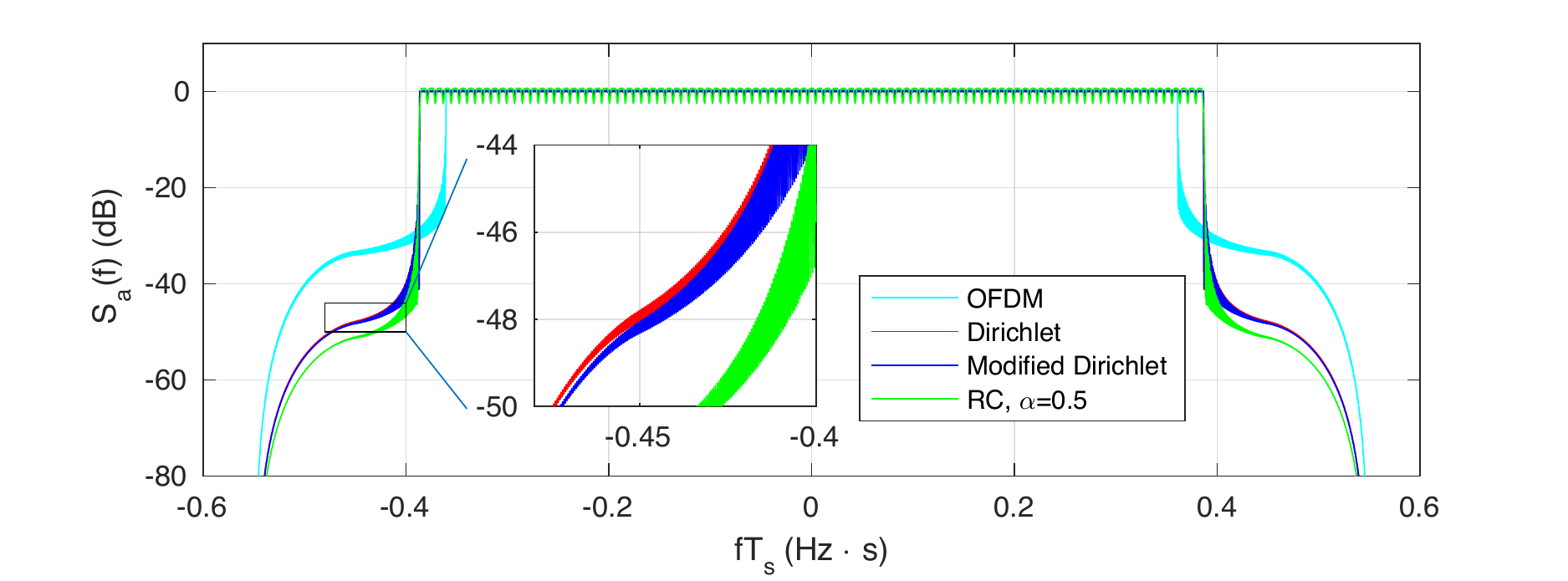}
\caption{PSD for GFDM and OFDM. The $0$th subsymbol is used as guard symbol, and subcarriers 50 to 78 are switched off, i.e., $\cK=\{0,1,\dots,49\} \cup \{79,80,\dots,127\}$, $\cM=\{1,2,\dots,14\}$.
Thus, $\cB_I=(-49.5,49.5)\cdot (1/(128T_s))$ Hz, $\cB_O=((-64(1+\alpha),-49.5-N_{gc}) \cup (49.5+N_{gc},64(1+\alpha)))\cdot (1/(128T_s))$ Hz, where $\alpha=0.1$, $N_{gc}=1$ or $6$.} \label{fig_sim_psd}
\end{figure}

Fig. \ref{fig_gf} shows the magnitude response of the prototype filters used for Case MMSE-RF, i.e., Fig. \subref*{MSE_MMSERx_RayleighCH_K8M4_SNR0to30}.
The Dirichlet and modified Dirichlet pulse{s} have the same magnitude response, and are more frequency-localized than the RC filter.
For Case MMSE-RF, we also evaluate the peak-to-average power ratio (PAPR) \cite{sharifian16}, defined as $\max |x[n]|^2 / \e\{|x[n]|^2\}$, where $x[n]$ is the digital baseband transmit signal (\ref{eq:dbts}).
PAPR complementary cumulative distribution function (CCDF) curves are shown in Fig. \ref{fig_papr}. The Dirichlet pulse, modified Dirichlet pulse, and OFDM are shown to have similar PAPR, while the RC filter has a higher PAPR.

\begin{table}[!t]
\renewcommand{\arraystretch}{1.3}
\caption{OOB leakage in dB of the simulation in Fig. \ref{fig_sim_psd}}
\label{tbl:oobgscp16}
\centering
\begin{tabular}{c|c|c|c|c}
\makecell{Guard\\carriers} & OFDM & \makecell{GFDM\\Dirichlet} & \makecell{GFDM\\modified Dirichlet} & \makecell{GFDM\\RC} \\
\hline
1 & -35.1 & -47.7 & -48.0 & -51.0 \\
\hline
6 & -37.1 & -51.5 & -51.8 & -54.8 \\
\end{tabular}
\end{table}

\subsection{OOB Leakage}
The PSD of GFDM and OFDM signals is simulated according to (\ref{eq:psd}), and the OOB leakage is evaluated according to (\ref{eq:oobdef}). The average in-band PSD is normalized to 1. 
We basically follow the simulation parameters in \cite{matthe14a}. 
We use $K=128$ and $M=15$ for GFDM. An RC filter with a roll-off factor of $0.5$, the Dirichlet pulse \cite{matthe14a}, and the modified Dirichlet pulse defined in (\ref{eq:designgl}) are used for GFDM, and the $0$th subsymbol is used as a guard symbol.
For a fair comparison, $K=1920$ and $M=1$ are used for OFDM so that the GFDM and OFDM block sizes are equal, and the number of used OFDM subcarriers is the same as the number of used resource elements in GFDM systems, $|\cK||\cM|$. Thus, the spectral efficiency of all systems with all filters are the same. The used OFDM subcarriers are contiguous, and their center is located at the DC bin.
The number of GFDM guard subcarriers used between $\cB_I$ and $\cB_O$ is $N_{gc}=1$ or $6$. The CP length is $L=16$. The interpolation filter $p(t)$ is a sample-level RC filter with roll-off factor $\alpha=0.1$, and the sampling rate is $1/T_s=1.92$ MHz.

We compare the OOB leakage of GFDM and OFDM systems, as presented in Table \ref{tbl:oobgscp16}, by using the simulated PSD shown in Fig. \ref{fig_sim_psd}. As shown in Table \ref{tbl:oobgscp16}, the Dirichlet pulse, which is optimal in terms of minimizing the MSE, outperforms OFDM by at least 12 dB, and has an OOB leakage comparable to that of the RC filter. Besides, the OOB leakage of the proposed modified Dirichlet pulse in this study is even lower than that of the Dirichlet pulse; this suggests that the Dirichlet pulse known in the literature is not optimal in terms of OOB leakage among all CMCM filters, and that we may further minimize the OOB leakage in the future with respect to the prototype transmit filter under the derived MMSE criterion.


\ignore
{
While Matthé et al. \cite{matthe14a} use the PSD defined for continuous-time signals, they do not specify the interpolation filter used for the D/A converter. If a sample-level sinc interpolation filter is used, then the PSD of the analog transmit signal is basically equivalent to that of the digital transmit signal. Since Fig. \ref{sfg:psd} and Tab. \ref{tbl:oobgs} match the results in \cite{matthe14a}, we guess that they indeed use a sample-level sinc interpolation filter.

}

\begin{figure}[!t]
\centering
\includegraphics[width=2.4in]{\figpath/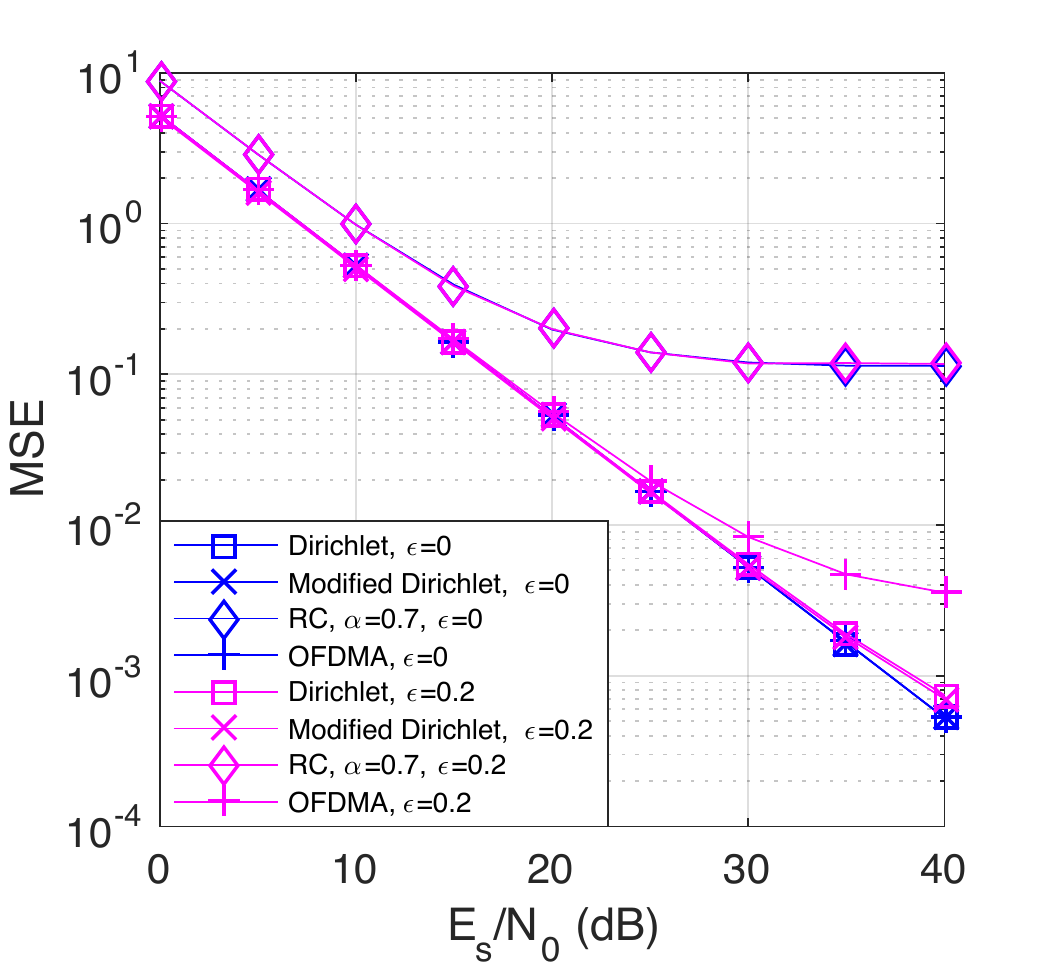}
\caption{MSE for GFDMA ZF receiver over a deep-fade-excluded Rayleigh fading channel and the corresponding OFDMA receiver. $K=32$, $M=15$.} \label{fig_sim_gfdma}
\end{figure}

\subsection{Multiple Access}
We briefly evaluate the performance of generalized frequency division multiple access (GFDMA) \cite{sharifian16} and compare it to that of orthogonal frequency division multiple access (OFDMA) \cite{lin10}.
In our simulation, uplink transmission is considered, and the same GFDM transmitter matrix with $K=32$ and $M=15$ is used by two users. The subcarriers used by the two users are $\cK=\{0, 1, \dots, 14\}$ and $\cK=\{16, 17, \dots, 30\}$, respectively, and the $0$th subsymbol is used as a guard symbol.
For a fair comparison, for OFDMA, we use $K=480$ and $M=1$, and the subcarriers used by the two users are $\cK=\{0, 1, \dots, 209\}$ and $\cK=\{240, 241, \dots, 449\}$, respectively, so that the spectral efficiency of GFDMA and OFDMA is the same.
We evaluate the performance of one user while assuming that the other user has a normalized carrier frequency offset \cite{lin10} $\epsilon=0$ or $0.2$ (normalized to the OFDM subcarrier spacing).
The ZF receiver under the same deep-fade-excluded Rayleigh fading channel as mentioned in Section \ref{ssc:mseser} is used.
The modulation is 16QAM, and the CP length is $L=D/4$.
Fig. \ref{fig_sim_gfdma} shows that the MSE performance of GFDMA using the Dirichlet pulse and modified Dirichlet pulse and OFDMA are the same when $\epsilon=0$.
However, when $\epsilon=0.2$, the Dirichlet pulse performs much better than OFDMA, and the modified Dirichlet pulse performs a little better than the Dirichlet pulse, which can be explained by their OOB leakage.
The RC filter performs the worst when $\epsilon=0$ and $0.2$ since its ZF prototype receiver filter is not frequency-localized and collects interference outside the desired bandwidth \cite{michailow14}.
The simulation result shows that GFDMA using the proposed CMCM filters is promising.

\section{Conclusions} \label{sec:conclusion}
A new matrix-based characterization of GFDM systems is proposed, which facilitates deriving properties of GFDM (transmitter) matrices not easily obtained under the traditional prototype-filter point of view. 
The class of unitary GFDM matrices is identified through the matrix characterization, and conditions for non-singularity of GFDM matrices can be expressed clearly with the new characterization.
Moreover, low-complexity transceiver implementations are derived on the basis of the characteristic matrix. 
Particularly, the necessary and sufficient conditions for the existence of a form of implementation with a linearithmic complexity for an MMSE receiver are derived. Such a receiver is determined to exist if the GFDM transmitter matrix is selected to be unitary. 
In the case where the implementation does not exist, a low-complexity suboptimal MMSE receiver is proposed, and its performance approximates that of an MMSE receiver, as shown by numerical results.
This study also reveals that prototype transmit filters minimizing the MSE under the ZF or MMSE receiver over various types of channels correspond to the class of CMCM filters, which subsequently correspond to scalar multiples of unitary GFDM matrices.
The simulation verifies the MSE optimality for the CMCM filters and shows that their SER performance is superior to that of non-CMCM filters.
Besides, the simulation indicates that the proposed modified Dirichlet pulse, which is a CMCM filter, has an OOB leakage that is lower than that of the Dirichlet pulse and comparable to that of the RC filter.
Finally, the advantage of GFDMA using the modified Dirichlet pulse over OFDMA is verified through numerical results.

In the future, a proof to Hypothesis 1, which states that CMCM filters minimize the MSE under the MMSE receiver over a statistical multipath channel, is desirable. Besides, the results in this paper suggest us to further study the trade-off between the OOB leakage and MSE or SER performance by designing the prototype transmit filter.
As a starting point before giving up MSE performance, we might design the phases of the entries in the characteristic matrix of a CMCM filter to obtain a minimum OOB leakage.
We might also design these phases to minimize PAPR.

\appendices

\vspace{-0.3cm}
\section{Proof of Theorem \ref{thm:mmsefs}} \label{adx:mmsefs}
\vspace{-0.1cm}
Using (\ref{eq:sigmakmv}) and noting that $\sigma^2 = \mathbf{1}_K^T {\boldsymbol\sigma} / K$, we obtain
\begin{IEEEeqnarray}{rCl}
    \sigma^2&=&\frac{N_0}{KD}\sum_{k=0}^{K-1}\sum_{l=0}^{M-1}\left(\sum_{r=0}^{K-1}\frac{1}{|C_{l+rM}|^2}\right)\frac{1}{|[\mG]_{k,l}|^2}.\label{eq:msefs} \IEEEeqnarraynumspace
\end{IEEEeqnarray}
Let $\alpha_{l}=\sum_{r=0}^{K-1}1/(|C_{l+rM}|^2)$. According to the Cauchy-Schwarz inequality, we have
\begin{IEEEeqnarray}{rCl}
    \left[\sum_{k,l}|[\mG]_{k,l}|^2\right]\left[\sum_{k,l}\frac{\alpha_{l}}{|[\mG]_{k,l}|^2}\right]\geq \left[\sum_{k=0}^{K-1}\sum_{l=0}^{M-1}\sqrt{\alpha_{l}}\right]^2,\label{eq:ineqfs} \IEEEeqnarraynumspace
\end{IEEEeqnarray}
where the equality holds if and only if $|[\mG]_{k,l}|^2/\sqrt{\alpha_l}$ is a constant in both $k$ and $l$. Using (\ref{eq:msefs}) and (\ref{eq:ineqfs}), we obtain
$
    \sigma^2\geq\frac{N_0}{KM^2\xi_G}(\sum_{l=0}^{M-1}\sqrt{\alpha_{l}})^2.
$

\vspace{-0.3cm}
\section{Proof of Theorem \ref{thm:mmseprob3}}
\vspace{-0.1cm}
\label{adx:mmseineq}
Let $\ve=\hat{\vd}-\vd$ be the error vector. By direct computation, we obtain that the autocorrelation matrix $\mR_e=\e\{\ve\ve^H\}$ is
\begin{equation}
    \mR_e=E_S(\mI_D - \mA^H(\mA\mA^H+\gamma^{-1}\mI_D)^{-1}\mA). \label{eq:autocor}
\end{equation}
Plugging (\ref{eq:cmimpl1}) into (\ref{eq:autocor}) yields $\mR_e=(\mW_M^H \otimes \mW_K) \mD_e (\mW_M \otimes \mW_K^H)$, where $\mD_e$ is a $D\times D$ diagonal matrix with $[\mD_e]_{k'+lK}=N_0/(|[\mG]|_{k',l}^2+\gamma^{-1})$, $\forall\; 0\leq k'< K, 0\leq l< M$. Since the magnitude of each entry of $(\mW_M^H \otimes \mW_K)$ is $1/\sqrt{D}$, we obtain $\forall\; 0\leq k< K, 0\leq m< M$,
\begin{IEEEeqnarray}{rCl}
    \sigma_{k,m}^2&=&[\mR_e]_{k+mK,k+mK}=\sum_{k'=0}^{K-1}\sum_{l=0}^{M-1}\frac{N_0/D}{|[\mG]|_{k',l}^2+1/\gamma}.\IEEEeqnarraynumspace \label{eq:unifmmseawgn}
\end{IEEEeqnarray}
According to the Cauchy-Schwarz inequality, we have $[\sum_{k',l}(|[\mG]|_{k',l}^2+\frac{1}{\gamma})^{-1}] [\sum_{k',l}(|[\mG]|_{k',l}^2+\frac{1}{\gamma})] \geq D^2$, where the equality holds if and only if $|[\mG]_{k',l}|$ is a constant in both $k'$ and $l$. Thus, $\sigma_{k,m}^2\geq E_S/(\gamma\xi_G+1), \forall\; 0\leq k< K, 0\leq m< M$, and the result follows from (\ref{eq:msedef}).

\vspace{-0.3cm}
\section{Proof of Corollary \ref{crl:uniform}} \label{adx:uniform}
\vspace{-0.1cm}
%
According to (\ref{eq:unifzfawgn}), we have $\sigma_{k,m}^2=\xi_HN_0$, $\forall\; 0\leq k<K$, $0\leq m<M$ under the ZF receiver over the AWGN channel. Thus, $\sigma_{k,m}^2$ is a constant in both $k$ and $m$ under this scenario.
The statement for the other two scenarios can be proved in a similar way by noting (\ref{eq:unifzfstt}) and (\ref{eq:unifmmseawgn}), respectively.

\vspace{-0.1cm}
\section*{Acknowledgment}
\vspace{-0.1cm}
The authors would like to thank the anonymous reviewers for their useful suggestions, which substantially helped to improve the quality of the paper.

\ifCLASSOPTIONcaptionsoff
  \newpage
\fi



\bibliographystyle{IEEEtran}
\bibliography{IEEEabrv,waveform}
%
%
%

\end{document}

%% file: newcommands.tex
\newcommand{\vc}{{\bf c}}
\newcommand{\vd}{{\bf d}}
\newcommand{\ve}{{\bf e}}

\newcommand{\vg}{{\bf g}}
\newcommand{\vh}{{\bf h}}

\newcommand{\vp}{{\bf p}}
\newcommand{\vq}{{\bf q}}
\newcommand{\vr}{{\bf r}}
\newcommand{\vs}{{\bf s}}
\newcommand{\vt}{{\bf t}}
\newcommand{\vu}{{\bf u}}
\newcommand{\vv}{{\bf v}}
\newcommand{\vw}{{\bf w}}
\newcommand{\vx}{{\bf x}}
\newcommand{\vy}{{\bf y}}
\newcommand{\vz}{{\bf z}}

\newcommand{\cA}{{\cal A}}
\newcommand{\cB}{{\cal B}}

\newcommand{\cK}{{\cal K}}

\newcommand{\cM}{{\cal M}}

\newcommand{\cX}{{\cal X}}

\newcommand{\bZ}{{\mathbb{Z}}}

\newenvironment{mat}[1]{\left[\begin{array}{#1}}{\end{array}\right]}

\newcommand{\mA}{{\bf A}}
\newcommand{\mB}{{\bf B}}
\newcommand{\mC}{{\bf C}}
\newcommand{\mD}{{\bf D}}
\newcommand{\mE}{{\bf E}}
\newcommand{\mF}{{\bf F}}
\newcommand{\mG}{{\bf G}}
\newcommand{\mH}{{\bf H}}
\newcommand{\mI}{{\bf I}}

\newcommand{\mM}{{\bf M}}

\newcommand{\mP}{{\bf P}}
\newcommand{\mQ}{{\bf Q}}
\newcommand{\mR}{{\bf R}}

\newcommand{\mU}{{\bf U}}
\newcommand{\mV}{{\bf V}}
\newcommand{\mW}{{\bf W}}

\newcommand{\ignore}[1]{}